\documentclass[twocolumn]{aastex63}

\usepackage{mathrsfs}

\accepted{to ApJ on 4 Nov 2023}

\usepackage[morefloats=10]{morefloats}

\shorttitle{galaxy quenching at the high redshift frontier}
\shortauthors{Bluck et al.}

\begin{document}

\title{Galaxy quenching at the high redshift frontier: \\A fundamental test of cosmological models in the early universe with JWST-CEERS}


\correspondingauthor{Asa F. L. Bluck}
\email{abluck@fiu.edu}

\author[0000-0001-6395-4504]{Asa F. L. Bluck}
\affiliation{Stocker AstroScience Center, Dept. of Physics, Florida International University,11200 SW 8th Street, Miami, 33199, Florida, USA}

\author[0000-0003-1949-7638]{Christopher J. Conselice} 
\affiliation{Jodrell Bank Centre for Astrophysics, Dept. of Physics and Astronomy, University of Manchester, Oxford Rd., Manchester M13\,9PL UK}

\author[0000-0003-2000-3420]{Katherine Ormerod} 
\affiliation{Jodrell Bank Centre for Astrophysics, Dept. of Physics and Astronomy, University of Manchester, Oxford Rd., Manchester M13\,9PL UK}

\author[0000-0003-1661-2338]{Joanna M. Piotrowska} 
\affiliation{Department of Astronomy, California Institute of Technology, 1200 East California Boulevard, Pasadena, California 91125, USA}

\author[0000-0003-4875-6272]{Nathan Adams}
\affiliation{Jodrell Bank Centre for Astrophysics, Dept. of Physics and Astronomy, University of Manchester, Oxford Rd., Manchester M13\,9PL UK}

\author[0000-0003-0519-9445]{Duncan Austin}
\affiliation{Jodrell Bank Centre for Astrophysics, Dept. of Physics and Astronomy, University of Manchester, Oxford Rd., Manchester M13\,9PL UK}

\author[0000-0002-6089-0768]{Joseph Caruana}
\affiliation{Department of Physics, University of Malta, Msida MSD 2080, Malta}
\affiliation{Institute of Space Sciences \& Astronomy, University of Malta, Msida MSD 2080, Malta}

\author[0000-0001-6889-8388]{K. J. Duncan}
\affiliation{Institute for Astronomy, University of Edinburgh, Royal Observatory, Blackford Hill, Edinburgh, EH9 3HJ, UK}

\author[0000-0003-1445-0590]{Leonardo Ferreira}
\affiliation{Centre for Astronomy and Particle Theory, University of Nottingham, Nottingham, UK}

\author{Paul Goubert}
\affiliation{Stocker AstroScience Center, Dept. of Physics, Florida International University,11200 SW 8th Street, Miami, 33199, Florida, USA}

\author[0000-0002-8649-2936]{Thomas Harvey}
\affiliation{Jodrell Bank Centre for Astrophysics, Dept. of Physics and Astronomy, University of Manchester, Oxford Rd., Manchester M13\,9PL UK}

\author{James Trussler}
\affiliation{Jodrell Bank Centre for Astrophysics, Dept. of Physics and Astronomy, University of Manchester, Oxford Rd., Manchester M13\,9PL UK}

\author[0000-0002-4985-3819]{Roberto Maiolino}
\affiliation{Kavli Institute for Cosmology, University of Cambridge, Madingley Road, Cambridge, CB3 0HA, UK}
\affiliation{Cavendish Laboratory—Astrophysics Group, University of Cambridge, 19 JJ Thomson Avenue, Cambridge, CB3 0HE, UK}


\begin{abstract}

\noindent We present an analysis of the quenching of star formation in massive galaxies ($M_* > 10^{9.5} M_\odot$) within the first 0.5 - 3\,Gyr of the Universe's history utilizing JWST-CEERS data. We utilize a combination of advanced statistical methods to accurately constrain the intrinsic dependence of quenching in a multi-dimensional and inter-correlated parameter space. Specifically, we apply Random Forest (RF) classification, area statistics, and a partial correlation analysis to the JWST-CEERS data. First, we identify the key testable predictions from two state-of-the-art cosmological simulations (IllustrisTNG \& EAGLE). Both simulations predict that quenching should be regulated by supermassive black hole mass in the early Universe. Furthermore, both simulations identify the stellar potential ($\phi_*$) as the optimal proxy for black hole mass in photometric data. In photometric observations, where we have no direct constraints on black hole masses, we find that the stellar potential is the most predictive parameter of massive galaxy quenching at all epochs from $z = 0 - 8$, exactly as predicted by simulations for this sample. The stellar potential outperforms stellar mass, galaxy size, galaxy density, and Sérsic index as a predictor of quiescence at all epochs probed in JWST-CEERS. Collectively, these results strongly imply a stable quenching mechanism operating throughout cosmic history, which is closely connected to the central gravitational potential in galaxies. This connection is explained in cosmological models via massive black holes forming and growing in deep potential wells, and subsequently quenching galaxies through a mix of ejective and preventative active galactic nucleus (AGN) feedback. 

\end{abstract}

\keywords{Galaxies: formation, evolution, star formation, quenching, feedback}


\section{Introduction}

\noindent The existence of quenched (quiescent, non-star forming) galaxies remains a significant puzzle in the field of extragalactic astrophysics. The reason for this is that galaxies cease forming stars long before the baryon content of their dark matter haloes are exhausted. Indeed, only $\sim$5-10\% of baryons collapse into stars by the present epoch (Fukagita \& Peebles 2004; Shull et al. 2011). Yet, simple models of galaxy formation (utilising only cosmological expansion, gravitation, and cooling processes) unanimously predict that the vast majority of baryons should have been collated into stars and stellar remnants by $z=0$ (e.g., Cole et al. 2000; Bower et al. 2006, 2008; Croton et al. 2006; Somerville \& Dav\'e 2015; Henriques et al. 2013, 2015). Closely related to this problem is the question as to why the majority of baryons in high mass galaxies, groups, and clusters reside in a hot gas halo ($T_{\rm Halo} \sim 10^{7-8}K$) which, despite being embedded in a $\sim$3(1 + z)\,K ambient medium (the Cosmic Microwave Background, CMB), is stable from cooling and collapse for {\it billions of years} (e.g., Fabian 1994, 1999, 2012; McNamara et al. 2000; Voit et al. 2002; McNamara \& Nulsen 2007; Hlavacek-Larrondo et al. 2012, 2015, 2018). 

Typically, theorists think about these problems in terms of cosmological star formation efficiency, i.e. the fraction of baryons converted into stars in a given dark matter halo (e.g., Henriques et al. 2015, 2019; Somerville \& Dav\'e 2015). The current best strategy to reduce the cosmological efficiency of star formation (and keep hot gas haloes from collapse) is to introduce baryonic feedback in two principal forms: (i) stellar and supernova feedback (Cole et al. 2000; Bower et al. 2008; Schaye et al. 2015); and (ii) active galactic nucleus (AGN) feedback (Croton et al. 2006; Sijacki et al. 2007, 2015; Bluck et al. 2011; Maiolino et al. 2012; Vogelsberger et al. 2014, 2015; Schaye et al. 2015; Zinger et al. 2020). The former is highly effective at slowing the rate of star formation in low mass systems, but fails to reduce star formation in the deep potential wells of massive galaxies (e.g., Fabian 2012; Henriques et al. 2019). On the other hand, AGN feedback becomes highly effective in high mass systems, which host more massive central black holes (e.g., Sijacki et al. 2007, 2015; Weinberger et al. 2017, 2018). Hence, contemporary theoretical models are in near consensus in their use, and requirement, of AGN feedback to explain the existence of massive quenched galaxies (e.g., Henriques et al. 2013, 2015; Vogelsberger et al. 2014, 2015; Schaye et al. 2015; Weinberger et al. 2017, 2018; Nelson et al. 2018; Pillepich et al. 2018; Dave et al. 2019; Zinger et al. 2020; Piotrowska et al. 2022).

Observationally, much skepticism surrounds the role of AGN in galaxy evolution, and especially their regulation of galaxy quenching. The primary reason for this is that a clear link between instantaneous AGN luminosity and reduction in star formation has not been forthcoming (see Bluck et al. 2023 for a discussion). Despite several early claims to support this link (e.g., Nandra et al. 2007; Bundy et al. 2008), the majority of contemporary observational studies of luminous AGN place these objects in actively star forming galaxies (e.g., Hickox et al. 2009; Aird et al. 2012; Rosario et al. 2013; Heckman \& Best 2014; Trump et al. 2015), apparently in contradiction to the notion of AGN feedback quenching galaxies. However, this objection (though logical) is the result of a fundamental misconception about the nature of AGN feedback in contemporary simulations, i.e. how AGN are actually theorized to quench galaxies.

As an example of this, in Piotrowska et al. (2022) we find that no strong link between specific star formation rate (sSFR) and current AGN accretion rate, or luminosity, is predicted by three state-of-the-art contemporary models (EAGLE, Illustris \& IllustrisTNG). This is in spite of the fact that all three of these simulations quench massive galaxies exclusively via AGN feedback. This result was confirmed in Ward et al. (2022), who find that the most powerful AGN reside in gas-rich star forming galaxies, the antithesis of quenching galaxies. Hence, it becomes of paramount importance to resolve this apparent discrepancy.

The answer lies in the distinction between power (i.e., AGN luminosity) and work done to the galactic system (i.e., some fraction of the energy emitted by AGN across cosmic time). Of course, energy is simply the time integral of power, so the two are fundamentally related, but they are by no means identical. Contemporary cosmological simulations unanimously predict that quenching should scale fundamentally with {\it energy} input from AGN over the lifetime of galaxies, {\it not} instantaneous power, luminosity, or identification of AGN (see Terrazas et al. 2020; Zinger et al. 2020; Piotrowska et al. 2022; Bluck et al. 2023). The goal for an effective observational test of the AGN feedback paradigm of massive galaxy quenching is thus to find an observable which is strongly correlated with AGN energy, not AGN luminosity. Fortunately, in essentially all physical AGN models, the energy released over the lifetime of a black hole is directly proportional to its mass (see Soltan 1982; Silk \& Rees 1998; Bluck et al. 2020a). 

Within Piotrowska et al. (2022), we find that all three of the simulations mentioned above predict that black hole mass should be the optimal observable predictor of massive galaxy quenching. Indeed, black hole mass is predicted to be far superior to dark matter halo mass, stellar mass, and black hole accretion rate (and hence AGN luminosity) in separating star forming and quiescent galaxies. In Bluck et al. (2023) we extend this analysis to moderately high redshifts. Across 10\, Gyr of cosmic history ($z = 0 - 2$), these simulations predict that the fundamental signature of AGN feedback should be a close dependence of quiescence on black hole mass. This is clearly identified via machine learning classification with Random Forests, RF (Piotrowska et al. 2022; Bluck et al. 2023). The distinct advantage of the RF approach is that it can be used to carefully disambiguate inter-correlated `nuisance' parameters from the fundamental {\it causal} relationships (see Bluck et al. 2022 for a thorough demonstration, where we show that this approach can reverse-engineer the `LGalaxies' semi-analytic model, Henriques et al. 2013, 2015).

In the local Universe, the key prediction of AGN feedback driven quenching has been tested in two complementary ways: (i) using a small sample of dynamically measured black hole masses (Terrazas et al. 2016, 2017; Piotrowska et al. 2022); and (ii) via the use of the black hole mass ($M_{BH}$) - central velocity dispersion ($\sigma$) relationship (e.g., Ferrarese \& Merritt 2000; Saglia et al. 2016), and other calibrations, to estimate black hole masses for samples of $\sim$500k galaxies (Bluck et al. 2016, 2020a,b, 2022; Piotrowska et al. 2022). The results from both tests are in close agreement with each other and, moreover, are precisely as predicted by AGN feedback models. 

Additionally, a fundamental link between quenching and the disordered (pressure-supported) kinematics in local galaxies is found in Brownson et al. (2022), with essentially no connection between quiescence and galactic rotation. This strongly suggests that quenching is critically linked to catastrophic redistribution of angular momentum within galaxies (from mergers or violent disk instabilities), which leads to both bulge and supermassive black hole growth (e.g., Hopkins et al. 2006, 2008). Hence, this kinematic result may be seen as yet further support for the AGN feedback paradigm in the local Universe.

At higher redshifts there is a dearth of dynamically measured black hole masses, and even measurements of central velocity dispersions are extremely rare. This is due to the fact that the vast majority of high-$z$ galaxy surveys are predominantly photometric, with limited (or no) spectroscopic counterparts. In the coming years this profound lack in our observational knowledge of early galaxies will be significantly remedied by the VLT-MOONRISE survey (Maiolino et al. 2020; Cirasuolo et al. 2020). However, in the meantime, we have found effective ways to test the AGN feedback paradigm in photometric data, through application of the virial theorem (see Bluck et al. 2022, 2023).

In Bluck et al. (2023), we find that cosmological models predict that, in lieu of black hole mass, the stellar gravitational potential (given approximately as the ratio of stellar mass to galaxy half-light radius, $\phi_* \sim M_* / R_h$) ought to become the best predictor of massive galaxy quiescence. Remarkably, we find that this prediction is precisely met across 10\,Gyr of cosmic history, through analyses of HST-CANDELS and the SDSS. Additionally, these comparisons with models provide a natural explanation to many prior observational studies which have found a close dependence of quenching on galaxy density, light concentration, and bulge mass (see, e.g., Wuyts et al. 2011; Fang et al. 2012; Cheung et al. 2013; Lang et al. 2014; Bluck et al. 2014, 2016, 2022).

With the launch of the James Webb Space Telescope\footnote{JWST Project: https://webb.nasa.gov/} (JWST, Gardner et al. 2006), and subsequent first data release in 2022 (Pontoppidan et al. 2022), we are now in a position to extend the test of the AGN feedback paradigm essentially to the very first quiescent systems in the Universe. The reason JWST is needed for this task is twofold. Firstly, the significantly larger JWST primary mirror, compared to HST, enhances the resolution of galaxies observed by (at least) a factor of three, where there is wavelength overlap. This is essential for accurate measurements of galaxy sizes at high redshifts. Secondly, and arguably even more importantly, the longer wavelength detection capabilities of JWST NIRCam (compared to HST ACS and NICMOS) provide a rest-frame optical (and near-IR) measurement of galaxy structure at the earliest cosmic epochs for the first time. This enables accurate constraints of the stellar mass distributions of early galaxies, since longer wavelengths are largely uncontaminated by bright young stellar populations (which typically contribute only a tiny fraction of the underlying mass budget of galaxies, yet dominate shorter wavelength measurements).

In this paper we apply techniques developed in Bluck et al. (2022, 2023), Piotrowska et al. (2022), and Brownson et al. (2022) to the largest public JWST galaxy survey (as of the time of writing): JWST-CEERS (Finkelstein et al. 2023; Kocevski et al. 2023; Kartaltepe et al. 2023). We also provide updated quenching predictions from two state-of-the-art simulations at the very early epochs we can now probe with JWST (EAGLE: Schaye et al. 2015; IllustrisTNG: Nelson et al., 2018; Pillepich et al. 2018), as well as consistent comparisons to lower-redshift galaxy surveys (i.e., HST-CANDELS: Grogin et al. 2011; Koekemoer et al. 2011, and the SDSS: Abazajian et al. 2009). This paper represents the first rigorous statistical test on the AGN feedback paradigm at cosmic dawn (the epoch in which the first galaxies form).

The paper is structured as follows. In Section 2 we discuss the various observational and simulated datasets used throughout this work. In Section 3 we describe our methodology for categorizing star forming and quenched galaxies throughout cosmic time. Additionally, we provide a brief summary of our machine learning classification technique utilising Random Forests (which is extensively discussed in prior publications). In Section 4 we present our results, split between theoretical predictions and observational tests, including the primary JWST-CEERS analysis. In Section 5 we discuss the results of this paper in the context of the theory of galaxy formation and evolution. We summarize our contributions in Section 6.


\section{Observations and Data}

\noindent In this paper we test the high-$z$ quenching predictions from hydrodynamical simulations using data from JWST-CEERS\footnote{JWST-CEERS Data Access: https://ceers.github.io/overview.html}  (Finkelstein et al. 2023; Kocevski et al. 2023; Kartaltepe et al. 2023; Bagley et al. 2023). We additionally compare to observations of later cosmic times in HST-CANDELS\footnote{CANDELS Data Access: https://archive.stsci.edu/hlsp/candels} (Grogin et al. 2011; Koekemoer et al. 2011), and the SDSS\footnote{SDSS Data Access: https://classic.sdss.org/dr7/access/} (Abazajian et al. 2009). Full details on the survey design, parameter measurements, and data access are provided in the above references. Here we give a review of the most important aspects of these data for this work.

\subsection{HST+JWST Data}

\noindent We utilize novel data from the public JWST-CEERS photometric survey (Finkelstein et al. 2023; Kocevski et al. 2023; Kartaltepe et al. 2023) to probe galaxy quenching at intermediate ($z = 0.5 - 2$), high ($z = 2 - 4$), and ultra-high ($z = 4 - 8$) redshifts. Our analysis of this data represents the first test of quenching mechanisms in the very early Universe, where we are just now discovering that AGN are likely very common (e.g., Juod{\v{z}}balis et al. 2023; Marshall et al. 2023; Ubler et al. 2023).  Furthermore, the JWST data, in combination with lower-$z$ surveys (see below), enable a unified analysis of star formation quenching in galaxies across essentially all of cosmic time. 

To analyze the structural properties of our galaxies, we utilize NIRCam JWST observations processed through a customized reduction process from the Cosmic Evolution Early Release Science Survey (CEERS; ID=1345). Specifically, we work with the sub-set of JWST data that overlaps with the Cosmic Assembly Near-IR Deep Extragalactic Legacy Survey (CANDELS; Grogin et al. 2011; Koekemoer et al. 2011) in the Extended Groth Strip field (EGS). The data is independently reduced using a custom setup of the \textsc{JWST} pipeline version \textsc{1.6.2}, with calibration files from \textsc{CDRS 0942}. For a detailed account of this process and the resulting data quality, refer to Ferreira et al. (2022).

Our parent sample comprises the same 3965 sources at redshifts $z > 1.5$ from the CANDELS catalogs, as described in Ferreira et al. (2022). At present, this represents one of the most extensive catalogs of galaxy classifications at $z > 1.5$ using JWST. To conduct our analysis, we rely on well-calibrated and reliable photometric redshifts, star formation rates, and stellar masses derived from the HST-CANDELS data (Duncan et al. 2014; Duncan et al. 2019; Whitney et al. 2021). The selection of sources does not involve morphological information, nor do we use magnitude cuts, ensuring the inclusion of sources that might appear faint in HST but are clearly visible in JWST observations. Additionally, we incorporate information on galaxies at lower redshifts within this field, processed in the same manner, and compare them with the comprehensive multi-field HST-CANDELS data at redshifts where galaxies are well enough resolved without JWST imaging (i.e., at $z < 2$, see Dimauro et al. 2018).

As explained in Ferreira et al. (2022), we adopt two distinct approaches to galaxy morphology and structure. First, we conduct visual classifications for all sources (see also Conselice et al. 2023, in. prep). Second, we employ quantitative structural measures using GALFIT (Peng et al. 2002), conducting parametric light profile fitting to measure sizes and Sérsic indices for each galaxy in each waveband available in the JWST observations (see Ormerod et al. 2023, in prep.). A comparison of visual morphologies with the quantitative structural measurements can be found in Ferreira et al. (2022). For this work, our focus primarily centers on the quantitative structures, which enables robust comparisons with cosmological models. However, the visual morphologies have been used in the JWST observations to test and validate the more sophisticated quantitative structural measurements.

\subsubsection{Photometric Redshifts, Stellar Masses\\ \& Star Formation Rates}

\noindent In this paper we utilize photometric redshifts in the Extended Groth Strip (EGS) field derived by Duncan et al. (2019). The software \textsc{eazy} (Brammer et al. 2008) is employed for template fitting, where three distinct template sets are used and fitted to the photometric bands. These templates incorporate zero-point offsets to modify input fluxes and address wavelength-dependent errors. In addition to these redshifts, a Gaussian process code (GPz; Almosallam et al. 2016) is employed to obtain further empirical estimates using a subset of the photometric bands. Individual redshift posteriors are calculated, and all measurements are combined in a statistical framework through a hierarchical Bayesian approach to arrive at a final redshift estimate. For a detailed explanation of the process, more details are available in section 2.4 of Duncan et al. (2019).

The data used to determine these redshifts are obtained from the original CANDELS+GOODS WFC3/ACS imaging and data, Spitzer/IRAC S-CANDELS (Ashby et al. 2015), and ground-based observations with CFHT (Stefanon et al. 2017). The methodology we use for this is described in detail in Duncan et al. (2019). The photometric redshifts are highly accurate for the redshift range studied in this paper, as validated when comparing with available spectroscopic redshifts within the EGS field (see Duncan et al. 2019 for various statistical tests).
 
To determine galaxy stellar masses, a modified version of the spectral energy distribution (SED) code (described in Duncan et al. 2014) is employed. This code allows for stellar mass measurement at all redshifts within the photo-$z$ fitting range. The stellar masses are also accompanied by a `template error function', described in Brammer et al. (2008), which accounts for uncertainties arising from the template set and any wavelength effects. Typical uncertainties on these measurements are estimated to be 0.2 - 0.3 dex.

More specifically, the stellar mass measurement technique employs Bruzual \& Charlot (2003) model spectral templates, encompassing a wide range of stellar population parameters, and we adopt and use a Chabrier 2003 initial mass function (IMF). The applied star formation histories follow exponential $\tau$-models for both positive and negative values of $\tau$, enabling the representation of both rising and declining star formation histories (which is crucial at high redshifts). The timescales utilized include: $\left|\tau\right|$ = 0.25, 0.5, 1, 2.5, 5 \& 10, along with a short burst model ($\tau$ = 0.05), and continuous star formation models.

To ensure the reliability of  our measured stellar masses, we compare these with other values independently calculated by other teams within the CANDELS collaboration (see Santini et al. 2015). This comparison aims to rule out systematic biases affecting the stellar mass estimates. While some scatter is observed between the two mass estimates ($\sigma \sim 0.3$ dex), the masses used in this study remain unaffected by any significant biases compared to others. Further details regarding the method and models can be found in Section 2.5 of Duncan et al. (2019).

The star formation rates (SFR) used in this work are derived from star formation histories extracted from the SED fitting of the galaxies (i.e., from the weighted sum of simple stellar populations of the best composite fit). From the best SED fit, the UV light from these galaxies is examined, and corrections are made for dust using the UV slope of the spectral energy distribution ($\beta$). This correction provides a measure of the dust attenuation in the galaxy, enabling the determination of the total star formation rate corrected for extinction. Our SFRs are found to agree extremely well with SFRs derived directly from the SED fitting (see Duncan et al. 2014).

\subsubsection{Galaxy Sizes from Surface Brightness Fitting}

\noindent Galaxy sizes are a crucial aspect of this paper. As such, we describe how these are measured here (although for full details see Ormerod et al. 2023, in prep.; Conselice et al. 2023, in prep.). 

We determine half-light radii and Sérsic indices for JWST-CEERS galaxies using the {\small GALFIT} software. {\small GALFIT} operates by fitting 2D Sérsic light profiles to each waveband image of each galaxy (Peng et al. 2002). Ultimately, {\small GALFIT} is a least squares fitting algorithm designed to find the optimum global solution to a 2D galaxy light profile, operating at its core a Levenberg-Marquardt algorithm. {\small GALFIT} determines the goodness of fit via the $\chi^{2}$ statistic for each free parameter configuration, and then systematically adjusts parameters until the $\chi^{2}$ in minimised. More precisely, {\small GALFIT} performs this optimization using the reduced chi-squared loss-function, $\chi_{\nu}^{2}$, determined by :

\begin{equation}
\label{eqn:gal_chi}
\chi_\nu^2=\frac{1}{N_{\mathrm{DOF}}} \sum_{x=1}^{n_x} \sum_{y=1}^{n_y} \frac{\left(f_{\mathrm{data}}(x, y)-f_{\mathrm{model}}(x, y)\right)^2}{\sigma(x, y)^2}
\end{equation}

\noindent which is summed over $n_x$ \& $n_y$ pixels, and where $N_{\rm DOF}$ is the number of degrees of freedom in the model. {\small GALFIT} uses a data image, $f_{data}(x,y)$, and a weight map, $\sigma(x,y)$, which are utilised together to calculate the best fit model image, $f_{model}(x,y)$ (see Peng et al. 2002). A full description of this method (along with numerous quality tests) is presented in Ormerod et al. (2023, in prep.).  

We utilize the obtained sizes and Sérsic indices for JWST-CEERS galaxies to explore the statistical connection between galaxy structures and the processes of star formation quenching at high redshifts. The JWST data brings a major advantage here, as we can now leverage these radius measurements to construct an accurate estimate of the stellar potential ($\phi_* \sim M_* / R_h$), and quantify the 2D and 3D stellar mass densities of galaxies within the CEERS field in the rest-frame optical. This is the first time that rest-frame optical structural measurements have been available at redshifts $z > 2$, thanks to the unique capabilities of JWST.\\

\subsection{Ancillary Observational Data I: HST-CANDELS}

\noindent We incorporate the full multi-field HST-CANDELS survey into our analysis to probe intermediate redshifts ($z = 0.5 - 2$). The limiting factor determining the maximum redshift is set by galaxy resolution and the need for rest-frame optical measurements (see, e.g., Dimaurao et al. 2018). As such, HST-CANDELS acts as a test on the intermediate redshift analysis from JWST-CEERS. The advantage is that the full HST-CANDELS survey is much wider, and hence probes a much larger volume at these epochs (with correspondingly higher galaxy counts).

Specifically, we utilize data from value-added catalogs\footnote{CANDELS VAC: https://mhuertascompany.weebly.com/data-releases- and-codes.html.} for the HST-CANDELS survey provided by Dimauro et al. (2018). From this data release we utilize photometric redshifts, stellar masses, star formation rates, and galaxy half light radii, as well as rest-frame optical magnitudes and colours (for various data quality checks). All of the above are constructed via a comprehensive morphological-SED pipeline, which also produces bulge \& disk masses, as well as a host of other structural information. The SED fitting is performed using {\small FAST} (Kriek et al. 2009), employing Bruzual \& Charlot (2003) stellar population synthesis models, and a Calzetti et al. (2000) extinction law. Full details on these procedures are provided in Dimauro et al. (2018).

\subsection{Ancillary Observational Data II: SDSS}

\noindent We also incorporate the Sloan Digital Sky Survey data release 7 (SDSS-DR7) legacy survey (Abazajian et al. 2009) into our analysis, to compare our high-$z$ results to the low-$z$ baseline. From the SDSS, we utilize a variety of public value added catalogs. We take spectroscopic redshifts from the parent data release (Abazajian et al. 2009). We take SFRs from Brinchmann et al. (2004). These are computed via two distinct modes: (i) for emission line, non-AGN galaxies, SFRs are estimated from dust corrected emission lines (particularly dependent on $H\alpha$); (ii) for non-emission line or AGN contaminated galaxies, SFRs are computed from an empirical relationship between sSFR and the strength of the 4000\,${\rm \AA}$ break (D4000). We take stellar masses from the SED catalogs of Mendel et al. (2014), and galaxy half light radii from the morphological catalogs of Simard et al. (2011). A concise summary of the techniques used in all of these measurements, as well as numerous data quantity checks, are provided in Bluck et al. (2014, 2016).

\subsection{Cosmological Simulations}

\noindent In this paper we extract the testable predictions for high-mass galaxy quenching within the first few Gyr of cosmic history from two state of the art cosmological hydrodynamical simulations: (i) EAGLE\footnote{EAGLE Data Access: http://icc.dur.ac.uk/Eagle/} (Crain et al. 2015; Schaye et al. 2015; McAlpine et al. 2016); and (ii) IllustrisTNG\footnote{IllustrisTNG Data Access: www.tng-project.org/} (hereafter TNG; Marinacci et al. 2018; Naiman et al. 2018; Nelson et al. 2018, 2019; Pillepich et al. 2018; Springel et al. 2018). In previous work we have also extracted quenching predictions from Illustris (Vogelsberger et al. 2014a,b). However, this latter simulation fails to predict quiescent objects in significant numbers at $z > 2$, so cannot be used at the epochs targeted in this work. Moreover, TNG replaces Illustris as a more realistic and successful version of the latter.

Full details on the simulations are given in the references above, as well a concise overview and comparison between them in Piotrowska et al. (2022). Additionally, a thorough description on the parameters used in this work are presented in Bluck et al. (2023). Here we extract the exact same parameters and apply the exact same quality checks as in our previous work, the only difference being that we now view the $z = 2, 3, 4$ snapshots, concatenated together, for our quenching analyses. A docker\footnote{Simulations Docker: https://hub.docker.com/u/jpiotrowska} describing how to extract these data from the simulation websites in provided in Piotrowska et al. (2022). The only modification needed is the snapshot redshifts (given above). 

In this work, for simulations, we select only high mass galaxies ($M_* > 10^{9.5} M_{\odot}$) in relatively high mass haloes ($M_{\rm Halo} > 10^{11} M_{\odot}$). This ensures that all galaxies studied are well resolved in the simulations. Furthermore, it enables robust comparison with high-$z$ JWST observations, which are essentially mass complete at these masses (e.g., Ferreira et al. 2022). We apply identical stellar mass cuts in observations, but cannot apply halo mass cuts since these are not available in the JWST and HST galaxy surveys. Nonetheless, we have checked that the halo mass criteria makes no significant difference to the simulation predictions. It is included primarily to exclude the possibility of studying galaxies without seed black holes (which are extremely rare at these stellar masses).

Given that these simulations are very well discussed in the literature, and that we have previously outlined our data extraction and quality assurance methodology, we give only a brief summary of the most relevant details for this work (i.e., with respect to galaxy quenching).

\subsubsection{EAGLE}

\noindent We utilize EAGLE-RefL0100N1504 (Schaye et al. 2015), which has a box size of $\sim 100 \, {\rm cMpc}^3$, and is run with the {\rm GADGET-3} smoothed particle hydrodynamics (SPH) code (Springel et al. 2005). The input cosmology is given by Planck Collaboration I (2014), assuming a spatially flat, dark energy - cold dark matter ($\Lambda$CDM) universe. The sole quenching mechanism for massive galaxies in EAGLE is AGN feedback. Supermassive black holes are seeded in low mass dark matter haloes and grow via Eddington-limited Bondi-Hoyle accretion (Hoyle \& Lyttleton 1939; Bondi \& Hoyle 1944). 

Feedback from AGN is modelled as a single mode in EAGLE. A fraction of the accreted rest energy of matter is converted to energy, with a fraction of that energy coupled to the interstellar medium (ISM). Energy exchange between the black hole accretion disc and the galaxy is modelled via a pure thermal injection, applied in a stochastic `bursty' manner (see Crain et al. 2015). This mechanism functions both to trigger quenching, via ISM outflows, and to keep massive galaxies quenched, by heating the surrounding circum-galactic medium (CGM). However, this model is known to be less effective at maintaining quiescence in high mass galaxies compared to observations (see Piotrowska et al. 2022).


\begin{figure*}
\begin{centering}
\includegraphics[width=0.8\textwidth]{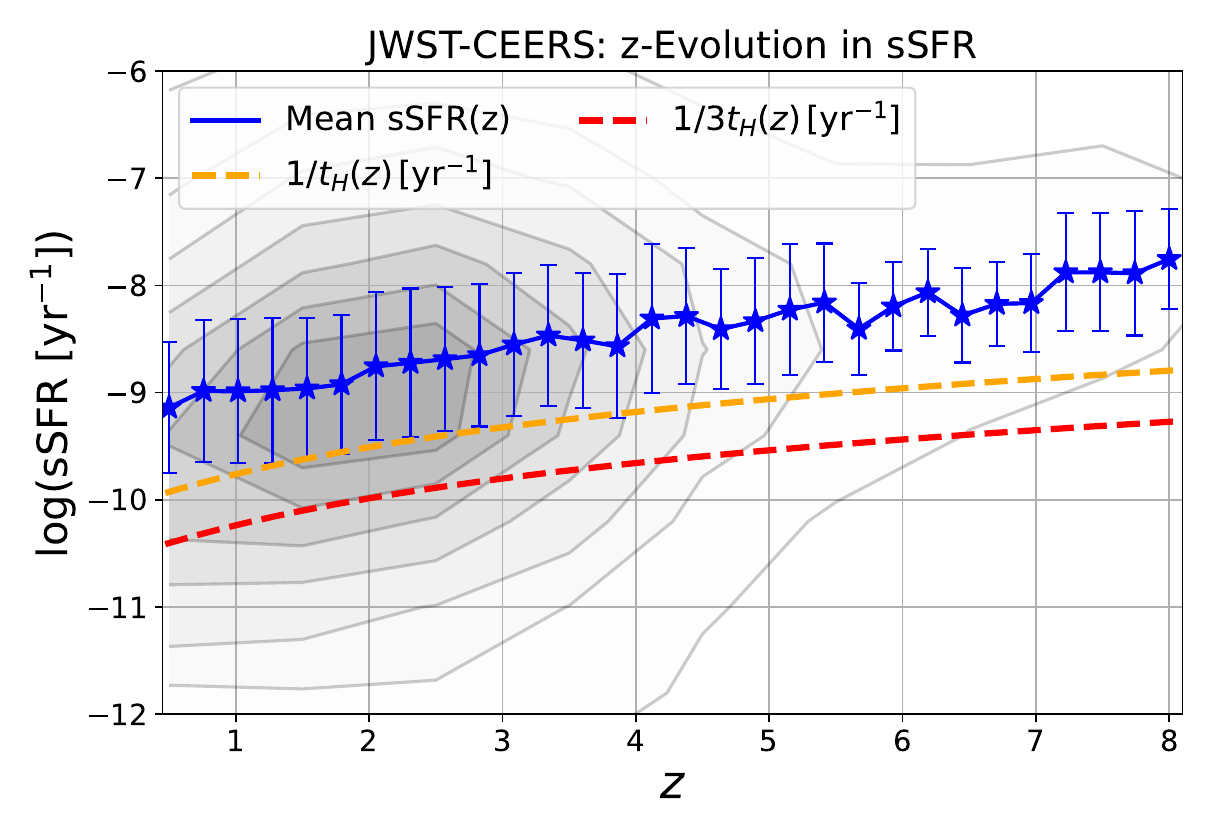}
\caption{Evolution in specific star formation rate (sSFR) as seen in JWST-CEERS. The mean $\log({\rm sSFR})$ at each epoch for star forming systems is presented by blue stars, with error bars displaying the full 1$\sigma$ range. The location of all galaxies in the sSFR - $z$ plane is displayed by grey contours (with linearly spaced intervals). Two quenching thresholds are displayed on the figure, representing: (i) $\mathrm{sSFR} < 1/t_{\rm{Hubble}}(z)$ (displayed in orange); and (ii) $\mathrm{sSFR} < 1/3t_{\rm{Hubble}}(z)$ (displayed in red). These thresholds clearly do a good job at all epochs studied in JWST-CEERS of separating star formers from quiescent systems. By design, the red threshold is more robust for identifying quiescence than the orange threshold, with the price of reduced numbers of quiescent systems. We use both definitions of quiescence throughout this paper for various purposes. Clearly, quiescent galaxies exist in the JWST-CEERS field up to $z = 6-7$. However, the vast majority of quenched galaxies reside at $z < 4$.}
\end{centering}
\end{figure*}

\subsubsection{IllustrisTNG}

\noindent We utilize TNG-100-1 (Nelson et al. 2018; Pillepich et al. 2018), which has a box size of $\sim 100 \, {\rm cMpc}^3$, and is run with the {\small AREPO} unstructured moving mesh code (Springel 2010), updated to implement magnetic fields in addition to gravity and hydrodynamics. The input cosmology is given by Planck Collaboration et al. (2016). As in EAGLE, supermassive black holes are seeded in low mass dark matter haloes and evolve in accordance with Eddington-limited Bondi-Hoyle accretion (Hoyle \& Lyttleton 1939; Bondi \& Hoyle 1944). 

Unlike in EAGLE, TNG operates three distinct forms of AGN feedback (see Sijacki et al. 2007; Nelson et al. 2018): (i) `quasar-mode' (high Eddington ratio); (ii) `kinetic-mode' (low Eddington ratio); and (iii) radiative ionization (all Eddington ratios). However, only the second mechanism (kinetic-mode) significantly impacts quenching in this simulation (see Weinberger et al. 2017; Zinger et al. 2020; Piotrowska et al. 2022). In this mode (which is only operational at $M_{BH} \gtrsim 10^8 M_{\odot}$), a fraction of the accreted rest energy of matter is converted to kinetic energy of gas cells in a stochastic manner (see Weinberger et al. 2017). This leads to turbulence and outflows in the ISM, as well as long term (preventative) heating of the CGM through gas percolation and shocks. The combination of these effects leads to quenching of massive galaxies in this simulation (see Zinger et al. 2020; Terrazas et al. 2020; and Piotrowska et al. 2022 for illuminating discussions). Note that the kinetic mode replaces the overzealous `radio-mode' in Illustris (see Sijacki et al. 2007), which had the undesired consequence of effectively destroying the CGM around massive galaxies.

\subsection{Sample Selection}

\begin{table}
\begin{center}
\caption{Data Summary}
\begin{tabular}{ l c c c } 
 \hline
Sample  & $N_{\rm gal}$ & $N_{SF}$ & $N_Q$    \\ 
  \hline
  \hline \\
{\bf OBSERVATIONS:} \\
JWST ($0.5 < z < 2$) & 384   & 243   & 141   \\ 
JWST ($2 < z < 4$) & 389   & 300   & 89   \\ 
JWST ($4 < z < 8$)$^*$ & 57   & 52   & 5    \\ 
CANDELS ($0.5 < z < 2$) & 9,417   & 6,034   &  3,383    \\ 
SDSS ($0.02 < z < 0.2$) & 498,206  & 247,364   & 250,842   \\\\

{\bf SIMULATIONS:}\\

EAGLE ($2 < z < 4$) & 5,999   & 5,289   & 710  \\
IllustrisTNG ($2 < z < 4$) & 10,890   & 10,011   & 879   \\\\
 \hline
\end{tabular}
\end{center}
Notes: All data is selected at $M_* > 10^{9.5} M_{\odot}$. Simulations are additionally required to have $M_{\rm Halo} > 10^{11} M_{\odot}$. The quenching threshold is set at sSFR = $1/H(z)$ (see Fig. 1), except for SDSS (which is set at sSFR = $10^{-11} {\rm yr}^{-1}$).\\ $^*$ Not used in machine learning analysis. {\bf Additionally note that the numbers of galaxies decrease rapidly with redshift, as a result of high-$z$ surveys having much smaller areas than low-$z$ surveys (see the source papers).}
\end{table}

\noindent For all machine learning analyses in this paper we select massive galaxies with $M_* > 10^{9.5} M_{\odot}$. For simulations, we additionally require that $M_{\rm Halo} > 10^{11} M_{\odot}$. In simulations we select only central galaxies (defined as being the most massive member in each dark matter halo) for extracting quenching predictions. This enables us to focus on intrinsic quenching mechanisms in simulations. For observations, we suspect that we are most sensitive to intrinsic quenching as well at these masses, but this is explicitly tested through comparison with the simulations (see also Bluck et al. 2023 for a discussion on this point).

We separate our sample into various redshift bins (see Table 1). The redshift cuts are motivated by studying galaxies at redshifts below and above cosmic noon ($z \sim 2$; see Madau \& Dickinson 2014) Additionally, we separate galaxies into star forming and quiescent systems based on their specific star formation rates (sSFR, see Section 3.1). The breakdown of total galaxy counts per sample, and the numbers of star forming and quiescent objects is provided in Table 1.

For the JWST observations we impose two quality cuts on the data: (i) {\it chi-square-robust = True}, which ensures high quality SED fits (see Duncan et al. 2014; 2019); and (ii) {\it Good-Fit = True}, which ensures high quality structural fits with {\small GALFIT} (see Ormerod et al. 2023, in prep.). For the SDSS and CANDELS data, we apply the exact same quality cuts as in Bluck et al. (2022, 2023).

To obtain the redshift range in simulations, we concatenate the $z = 2, 3, 4$ snapshots from both EAGLE and TNG. We have also checked that our analysis is stable to viewing each snapshot individually.

Towards the end of Section 4 we extend our analysis to explore galaxies at lower masses in the JWST data ($M_* > 10^{9} M_{\odot}$), where we are sensitive to environmental quenching routes in addition to intrinsic quenching.

\newpage


\section{Methods}

\subsection{Identifying Quiescent Galaxies across Cosmic Time}

\noindent In Fig.~1 we present the evolution in specific star formation rate (sSFR $\equiv {\rm SFR} / M_*$) as a function of redshift for galaxies observed in JWST-CEERS. Grey contours show the location of all galaxies. The geometric mean sSFR for galaxies with sSFR $> 10^{-10} \, {\rm yr}^{-1}$ (the quenched threshold at $z = 0.5$, see Bluck et al. 2023) is presented by blue stars, with error bars indicating the full 1\,$\sigma$ range. It is clear that the typical sSFR of star forming galaxies rises significantly with redshift, all the way out to $z=8$, i.e. to $\sim$650 Myr after the Big Bang. This justifies the cut in sSFR above, which is used as a coarse method to remove obviously quenched systems before assessing evolution in star formation, and ultimately constructing a more accurate quenching threshold (discussed below).

From the density contours in Fig.~1, it is clear that there is a significant population of galaxies with sSFR much lower (and higher) than the median relation. To identify quiescent systems we utilize the common approach at intermediate-to-high redshifts of considering the inverse Hubble time (e.g., Tacchella et al. 2019; Bluck et al. 2022, 2023). This is displayed on Fig. 1 as a dashed orange line, which clearly does a good job of identifying low sSFR systems (relative to the norm of the star forming population). Additionally, we include a more stringent cut at a value of sSFR a factor of three lower (shown as a red dashed line). Using these thresholds, we see in Fig. 1 that quiescent galaxies exist up to $z = 6 -7$ (as confirmed in, e.g., Looser et al. 2023), yet the vast majority of quenched galaxies in JWST-CEERS are at $z < 4$. 

Throughout this paper we define quenched galaxies to be any system which has an sSFR below the inverse Hubble time (at its epoch in cosmic history), with star forming systems being any galaxy with sSFR above this threshold. Additionally, we define fully quenched systems to be galaxies with sSFR values below a third of the inverse Hubble time (as in Tacchella et al. 2019). All results remain unchanged for either quenching threshold, indicating that our results are highly stable to the exact threshold adopted for separating star forming and quiescent objects. We apply the exact same quenching thresholds to all observational and simulated datasets used in this work to ensure a fair comparison (with the sole exception of the $z=0$, SDSS data, which is discussed below).

Utilising data from HST-CANDELS, we have found that the sSFR quenching threshold approach leads to extremely similar quiescent populations to $UVJ$ color diagnostics, as well as to a simple -1\,dex offset from the peak sSFR at each epoch (see Bluck et al. 2022, 2023). This indicates that the choice of method used to identify quiescent systems is not a significant source of bias or uncertainty at $z < 2$. However, at very high redshifts galaxy color is more problematic as a quenching indicator because essentially all stellar populations are relatively young (due simply to the age of the Universe). As such, an sSFR approach calibrated on finding evidence of very massive (short lived) stars is a superior approach (see, e.g., Carnall et al. 2023)

At very low redshifts (i.e. in the local Universe), the above defined quenching thresholds break down. In fact, the star forming main sequence relationship perfectly intersects with the inverse Hubble time at $z = 0$, making it clearly invalid as a quenching definition. The reason for this is likely due to dark energy leading to accelerated expansion, and hence exponential decline in gas accretion (see Henriques et al. 2013, 2015). As such, for the SDSS data (which we use as a low-$z$ comparison to our high-$z$ results), we adopt the standard $z = 0$ quenching threshold of sSFR $< 10^{-11} \, {\rm yr}^{-1}$ (see, e.g., Piotrowska et al. 2022; Bluck et al. 2022, 2023). This yields a similar offset from the mean sSFR of star forming systems as the inverse Hubble time definition at higher redshifts (see Bluck et al. 2022).

One potential issue with the sSFR approach is that the star forming main sequence (${\rm SFR} - M_*$) relation is actually slightly sub-linear (with a gradient of 0.7 - 0.8; see, e.g., Bluck et al. 2016, 2020b), whereas a threshold in sSFR implicitly assumes a gradient of unity. As such, a fixed threshold in sSFR slightly over identifies quiescent galaxies at very high masses. However, this effect is very small and we have checked that alternative $\Delta MS$ methods (i.e. offsets from a main sequence fit) yield equivalent results to the more straightforward sSFR criterion.

Ultimately, the main problem with measuring offsets from the star forming main sequence is that this relation is quantitatively different in simulations and observations, in addition to its intrinsic redshift evolution. Conversely, the sSFR method is an objective definition of quenching, which can be used in simulations and observations without the need for fitting bespoke star forming main sequence relations (which may obscure important differences between models and observations). As such, we choose the sSFR method for its objectivity and reproducibility (see Bluck et al. 2023 for further discussion on this point). Nonetheless, we note that this yields identical conclusions to a $\Delta {\rm MS}$ approach in any case (see also Bluck et al. 2022 for a demonstration of this).

\subsection{Random Forest Classification}

\noindent In Bluck et al. (2022) we present a detailed overview of Random Forest (RF) classification, including a series of tests which demonstrate its capacity to accurately distinguish causal from nuisance parameters in simulated data. Ultimately, this is the key advantage of this machine learning technique over other approaches. The RF approach has been used in numerous previous papers to hone in on the fundamental drivers of galaxy quiescence (see Bluck et al. 2020a,b; 2022; 2023; Piotrowska et al. 2022; Brownson et al. 2022). As such, we give only a brief outline of the most salient aspects of this method here. 

RF classification is a type of supervised machine learning which provides an effective route to categorize data into types, based on given input features (or parameters). It is capable of discerning highly non-linear boundaries in a complex multi-dimensional parameter space (see Bluck et al. 2022 for several demonstrations). However, unlike deep learning with, e.g., artificial neural networks (ANN), the RF approach is emphatically not a `black box'. Every step in the iteration may be output and, given the simple criteria for developing each tree (discussed below), clearly understood by humans\footnote{However, note that strategies exist to achieve this with deep learning as well, via the use of interpreters. Yet, such methods are highly complex and infrequently incorporated in standard machine learning.}. Consequently, this approach yields an optimal compromise between sophistication and interpretability, which is ideal for the application to astronomical data. The power of this approach has been thoroughly demonstrated by its capacity to accurately reverse-engineer both semi-analytic and hydrodynamical simulations, i.e. revealing the input physics in these models from their output galaxy catalogs (see Bluck et al. 2022, 2023; Piotrowska et al. 2022).

Briefly, an RF is a set of decision trees with differences enforced from bootstrapped random sampling. Within a given tree, the classifier chooses the most effective feature and threshold to minimise the impurity (quantified by the Gini parameter) of a sample. This continues throughout each level of the decision tree until all data is perfectly segregated, or else a pre-defined limit is reached. This limit can be used to mitigate the potential for overfitting (see below). Finally, the `votes' on the class of each object in the sample are averaged over each tree in the forest to yield a final class prediction and a probability distribution, which encodes the uncertainty of that prediction.

In this work we construct an RF classifier using the {\small SCIKIT-LEARN} {\small PYTHON} package (Pedregosa et al. 2011). Following our exploration in Bluck et al. (2022), we make every feature available to the classifier at every step in the RF, which enables disambiguation of inter-correlated nuisance parameters up to correlations of $\rho = 0.99$, or higher (see Piotrowska et al. 2022). As such, our bespoke RF approach offers one route to move beyond the canonical problem of correlation not implying causation. By using RF classification one can definitively prove that a parameter is only useful in a classification problem due to its inter-correlation with another parameter (i.e. that it is non-causal). An interesting case in point is that of stellar mass, which whilst well known to correlate strongly with quiescence (see Baldry et al. 2006; Peng et al. 2010, 2012), is clearly established to be non-causal via RF classification in the local Universe (see Bluck et al. 2020a, 2022, 2023; Piotrowska et al. 2022; Brownson et al. 2022).

As in all supervised machine learning, it is always possible to perfectly solve the problem on test data (seen by the classifier). However, the real value of the RF classification is that it can be effectively applied to novel data (unseen by the classifier). To ensure this, we split the sample into training (typically 50 - 70\,\% of the data) and testing (typically 30 - 50\,\% of the data), and require that the performance (as measured by the area under the true positive - false positive receiver operator curve, AUC, see Teimoorinia et al. 2016) is extremely similar in both samples. Specifically, we require $|\Delta {\rm AUC}| < 0.02$ (see Bluck et al. 2023). This ensures very similar performance on unseen samples to the training data, which prevents the classifier from learning pathological features in the data. Consequently, the RF classifier learns to classify novel data almost as accurately as its training sample.

Due to the transparency of the RF architecture, the importance of a given feature for solving the classification problem can be extracted. It is given simply as the weighted average across the entire RF of the improvements in Gini coefficients engendered by each feature in turn, normalized by the total reduction in Gini engendered by all parameters (see Bluck et al. 2022, 2023). As such, this gives a clear intuitive sense of how valuable (or important) a given parameter is for solving the classification problem, within the context of the chosen set of parameters. Consequently, it is a {\it relative importance}: an importance relative to the available list of parameters. As such, one can always rule out causality in a given parameter for a given application (e.g., quenching), but can only find causality in the limit where every conceivable parameter has been assessed. Hence, one approaches causality from RF classification via a process of elimination.

In this work we use RF classification to separate star forming and quiescent galaxies in observations and simulations (based on the training thresholds discussed in Section 3.1). We train the RF classifier with a variety of parameters, which are known or suspected to be connected with various quenching mechanisms. The RF classifier then sorts the parameters by how valuable they are for predicting quiescence in each dataset. We then use these results to test theoretical models, and ultimately identify the underlying causes of transition from star formation to quiescence in galaxies across cosmic time. We also provide other simpler statistical tests on the main RF classification results to establish confidence in these relatively new methods. Readers interested in the full mathematical details of RF classification are encouraged to read Appendix B in  Bluck et al. (2022). Additionally, in Appendix A of this work, we provide full details on the RF hyperparameters used for each classification analysis in this paper. This is to ensure reproducibility of our results.


\section{Results}

\subsection{High-z Quenching Predictions from Hydrodynamical Simulations}


\begin{figure*}
\begin{centering}
\includegraphics[width=0.7\textwidth]{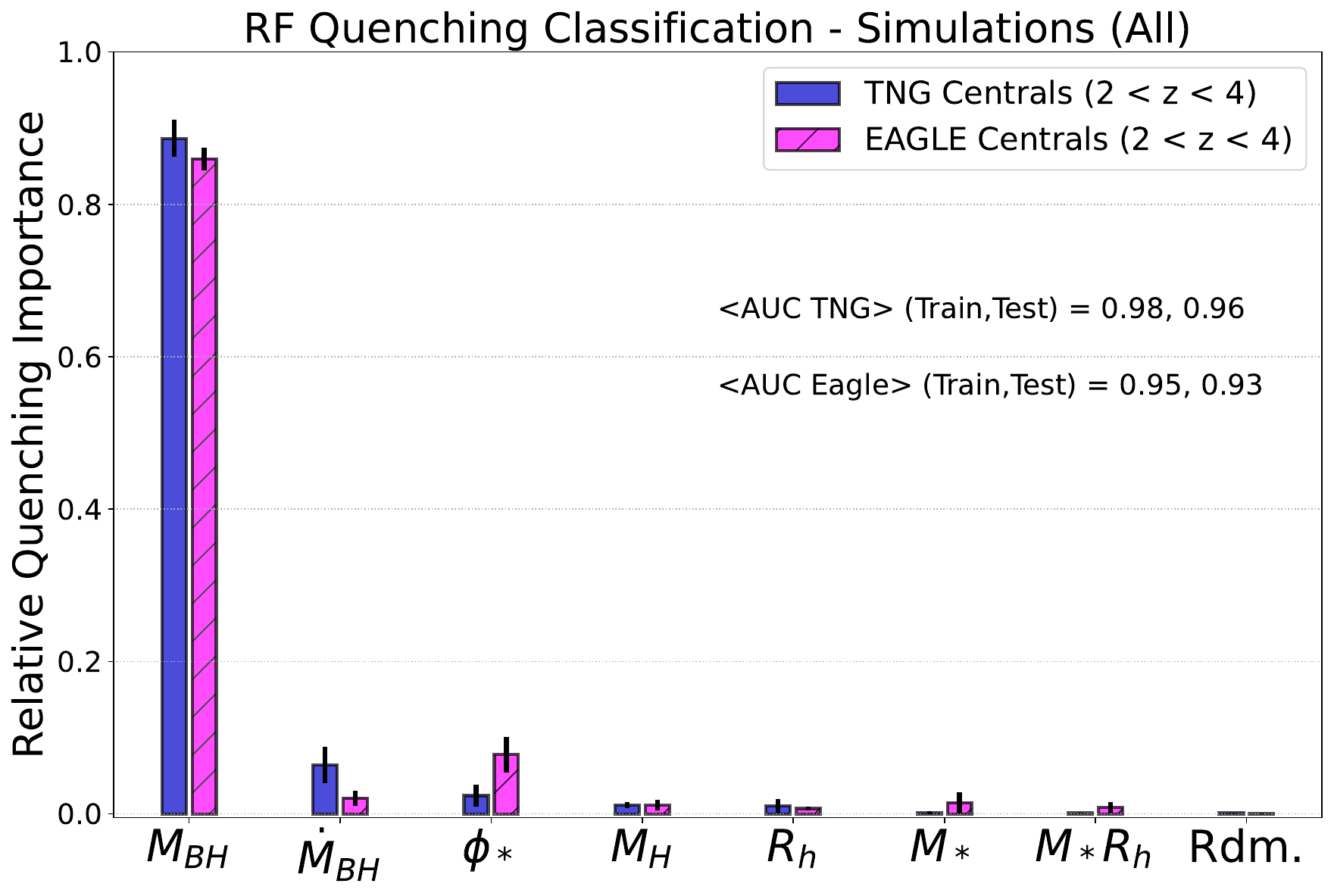} 
\includegraphics[width=0.7\textwidth]{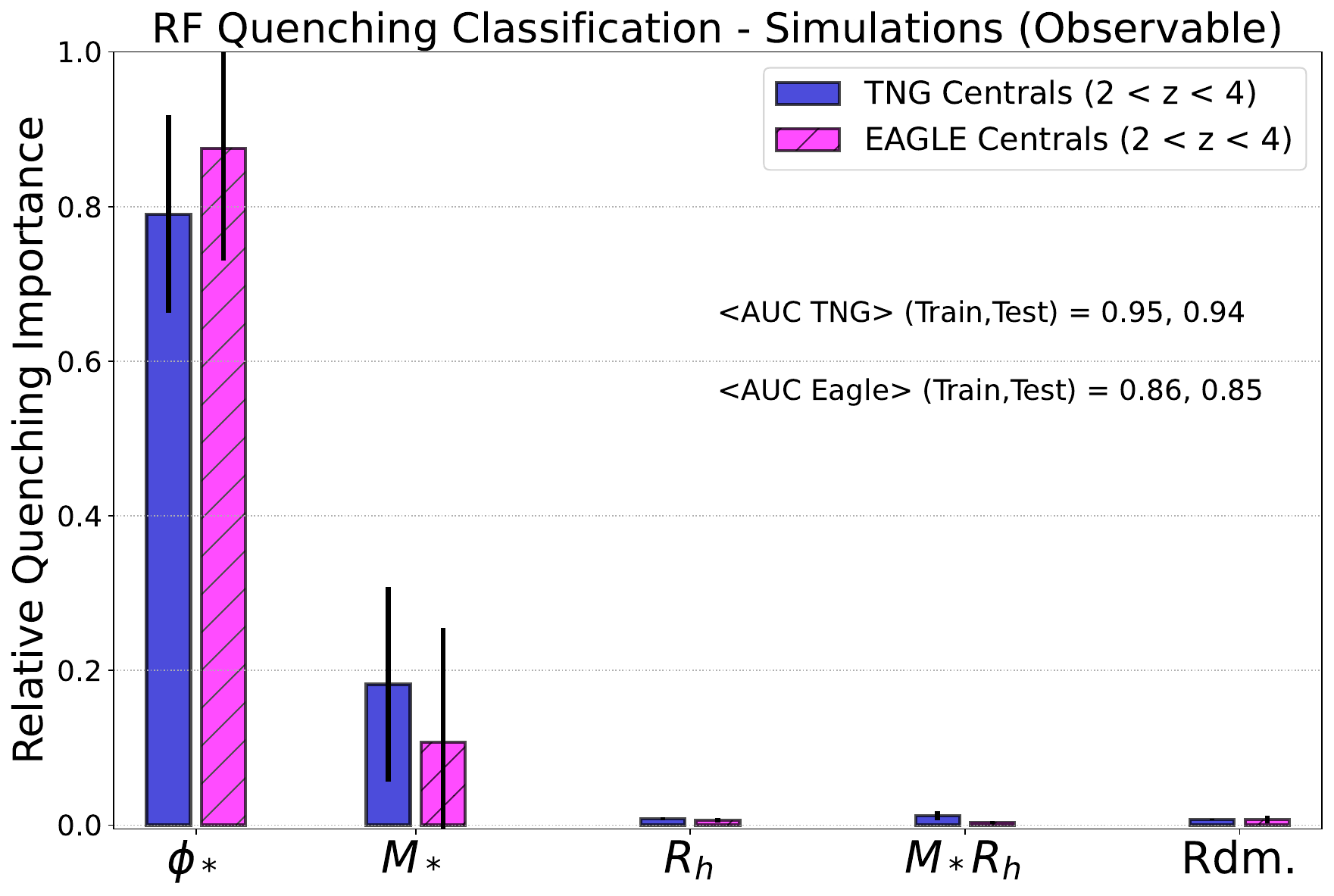}
\caption{Random Forest quenching classification analyses to predict when galaxies will be star forming or quiescent in cosmological simulations at high redshifts. Results are shown separately for the TNG and Eagle cosmological simulations (see legends). The top panel displays results for training with all parameters ($M_{BH}$, $\dot{M}_{BH}$, $\phi_*$, $M_{H}$, $R_{h}$, $M_*$, $M_* R_h$, Rdm.). The bottom panel displays results for a reduced set of parameters, which are possible to measure accurately in extant observational data at high-z from JWST. The relative importance for quenching is displayed on the $y$-axis for each parameter. Uncertainties on the quenching importance are inferred from the variance over multiple RF classification runs. For the full parameter set (top panel), it is evident that supermassive black hole mass is clearly the critical parameter for setting quenching in the models. Interestingly, there is essentially no importance given to black hole accretion rate, dark matter halo mass, stellar mass, or galaxy stellar potential, once black hole mass is controlled for. However, in the restricted set of parameters (bottom panel), we see that $\phi_*$ is predicted to be found as the best observable parameter for quenching in lieu of black hole mass. This is a key testable {\it prediction} of AGN feedback quenching in the early universe for photometric surveys. See Bluck et al. (2023) figs. 2 \& 3 for equivalent results at lower redshifts. See Table 2 in Appendix A for full details on the RF hyperparameter setup.}
\end{centering}
\end{figure*}

\noindent In Bluck et al. (2023) we expand on the low-$z$ analysis of Piotrowska et al. (2022) to reveal the key quenching predictions of hydrodynamical simulations at $z = 0 - 2$. We find that in EAGLE, Illustris, and IllustrisTNG (hereafter, TNG), black hole mass ($M_{BH}$) is the clear best parameter for predicting quiescence. This is explained as a consequence of the total energy released by AGN feedback over the lifetime of a galaxy being directly proportional to the mass of the central black hole (see Bluck et al. 2020a). In lieu of a dynamical measurement of black hole mass, we find that the stellar potential ($\phi_*$) is predicted by these simulations to act as a excellent proxy, outperforming many other parameters in predicting quiescence in models (Bluck et al. 2023). 

In this sub-section, we extend the AGN feedback predictions from EAGLE and TNG (Illustris has essentially no quiescent galaxies at $z>2$) to the highest redshifts at which quiescent galaxies exist in these simulations.

In Fig.~2 we show results from a set of RF classification runs used to separate central galaxies (at $M_* > 10^{9.5} M_{\odot}$) into star forming and quenched categories, based on the threshold: sSFR = $1/t_{H}(z)$ (orange dashed line in Fig. 1). We have also tested the impact of using the more severe cut in quenching (sSFR = $1/3t_{H}(z)$, red dashed line in Fig.~1) and find that the following conclusions remain completely stable. We use central galaxies in the simulations to extract the quenching predictions for AGN feedback in isolation (since only satellites experience strong environmental quenching in these models).

In the top-panel of Fig.~2, we show results from using a large variety of parameters to train the RF classifier: black hole mass ($M_{BH}$); black hole accretion rate ($\dot{M}_{BH}$); stellar gravitational potential ($\phi_* \equiv M_*/R_H$); dark matter halo mass ($M_H$); galaxy half-mass radius ($R_h$); stellar mass ($M_*$); mass times radius ($M_* R_h$); and finally a random number (Rdm.), for comparison purposes. These parameters have been carefully chosen to represent various potential quenching channels (e.g., supernova feedback - stellar mass; virial shocks - halo mass; AGN feedback - black hole mass or black hole accretion rate, depending on the mode of operation, etc.). 

In the top-panel of Fig. 2 we see that black hole mass is overwhelmingly the most valuable parameter for identifying quenched systems, vastly outperforming all other variables in both the EAGLE and TNG simulations at very high redshifts ($2 < z < 4$). Note that this is the first time we have determined the key observable predictors of quiescence in these simulations at these very early cosmic times (but see Piotrowska et al. 2022; Bluck et al. 2023 for the predictions at later epochs, $z < 2$). This is essential for accurate comparison to the JWST observations (see Section 4.2). At even higher redshifts, the number of quiescent galaxies falls to zero in TNG and near to zero in EAGLE, so further statistical analysis at even earlier cosmic times is not possible in these simulations.

Note that the lack of importance between stellar mass, halo mass, stellar potential and quiescence in Fig. 2 (upper panel) does not indicate a lack of correlation. In the Random Forest, all other parameters are controlled for when assessing the importance of any given training parameter. Hence, the lack of importance indicates a lack of residual information relevant to quenching once black hole mass is made available to the classifier. This clearly establishes that the well-known correlations between quenching and these parameters are ultimately spurious, i.e. of non-causal origin, originating solely from inter-correlation with the causal black hole mass.

It is especially important to highlight the lack of importance of accretion rate in models for predicting galaxy quiescence in the very early Universe. Hence, efforts to test AGN feedback in the early Universe by using tracers of accretion rate (i.e., AGN luminosity or detection fractions) are destined to failure. The correct approach to test the modern AGN feedback paradigm is to look for parameters closely connected with black hole mass, and hence the integrated energy released from AGN over the lifetimes of their host galaxies (see Section 1; Piotrowska et al. 2022; Bluck et al. 2023 for further discussion).

Given that extremely few black hole masses have been dynamically measured in the very early Universe (and none in our sample), we must now look for a predicted proxy in order to test the hydrodynamical predictions in observations. Following the logic of Bluck et al. (2023), there is strong evidence to suspect that supermassive black hole mass should scale strongly with the gravitational potential of the galaxy (a consequence of the virial theorem in combination with the $M_{BH} - \sigma$ relation). Here we estimate the gravitational potential with the stellar gravitational potential, which we take to be approximately given by the ratio of stellar mass to galaxy half-mass radius (see Bluck et al. 2023 for many caveats). See Appendix B of this paper for an explicit demonstration of the strong relationship between $M_{BH}$ and $\phi_*$ predicted by both TNG and EAGLE at early cosmic times. 

In the bottom-panel of Fig.~2, we restrict the number of variables used to train the RF classifier to those which can be accurately constrained in JWST data. In this reduced parameter set, both Eagle and TNG predict that quenching should be best constrained by $\phi_*$, which significantly outperforms $M_*$, $R_h$, and $M_* R_h$. The latter variables all perform (at best) only marginally better than a random number, once $\phi_*$ is available to the classifier. This result is expected from the simple arguments sketched above (and in Bluck et al. 2023). Hence, the simulations predict that black hole quenching will look like stellar potential quenching in photometric data.

Taken as a whole, Fig.~2 shows that (in both EAGLE and TNG) $M_{BH}$ is the fundamental parameter which regulates central galaxy quenching at early cosmic times. Moreover, in observational data, AGN feedback driven quenching should look like a strong dependence of quiescence on $\phi_*$. This represents a clear testable prediction of the AGN feedback paradigm in the very early Universe. 

It is important to note the stability of this prediction to the specifics of the AGN quenching model (which are very different in Eagle and TNG; compare Schaye et al. 2015 with Weinberger et al. 2017, 2018). The reason for the similarity in their predictions is a consequence of both models ultimately extracting energy from around supermassive black holes to quench massive galaxies. The exact nature of the extraction, its coupling to the galaxy and halo, and the triggering mechanism(s) are all secondary to the crucial fact of the energy source. Consequently, there is a clear way to test the entire paradigm of AGN feedback, not just one specific instantiation of it (see Bluck et al. 2020a, 2023 for more details on this important point). In the next sub-section we perform this test in JWST-CEERS.\\


\begin{figure*}
\begin{centering}
\includegraphics[width=0.8\textwidth]{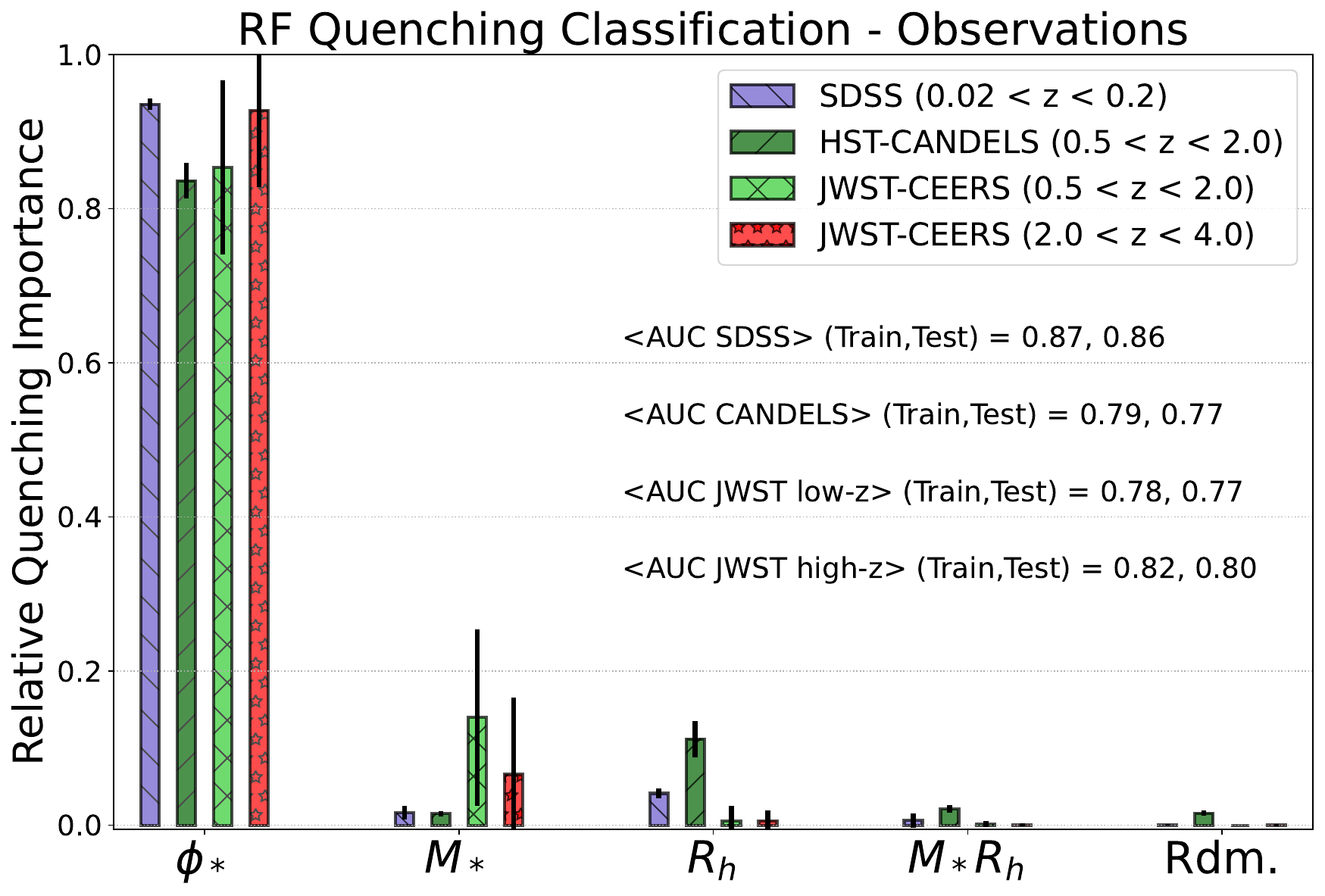}
\caption{Random Forest quenching classification comparing observational datasets targeting galaxies across $\sim$13\,Gyr of cosmic history ($z = 0 - 4$). The $x$-axis displays the parameters used to train the RF classifier, ordered from most to least predictive of quiescence (averaged across all epochs). The $y$-axis displays the importance of each variable for solving the classification problem at each epoch (see the legend). Uncertainties on the quenching importance are given from the variance of multiple independent RF classification runs, randomly sampling the data in each galaxy survey. At all epochs, the stellar potential ($\phi_*$) is found to be overwhelmingly the most important parameter for predicting quiescence in galaxies. Stellar mass ($M_*$), galaxy half-light radius ($R_h)$, and mass times radius ($M_* R_h$) are all either equivalent in importance to a random variable (Rdm.), or else only marginally superior to it. This result clearly implies a stable quenching mechanism operating across cosmic time, which is closely linked to the galaxy gravitational potential. Moreover, this result is precisely as predicted by contemporary cosmological models utilizing AGN feedback to quench galaxies (compare to Fig.~2 lower panel, and Bluck et al. 2023 for lower-$z$ results). See Table~3 in Appendix A for full details on the RF hyperparameter setup.}
\end{centering}
\end{figure*}

\subsection{HST+JWST Results}

\subsubsection{Random Forest Classification}

\noindent In Fig.~3 we present results from an RF classification to predict quiescence in galaxies, applied to JWST-CEERS (at $z=0.5-2$ and $z=2-4$). For comparison, we show equivalent results from HST-CANDELS (at $z=0.5-2$), and the SDSS (at $z=0.02-0.2$). The parameters used in training of the classifier are listed along the $x$-axis, and are identical to those used in the bottom-panel of Fig.~2 for simulations. The quenching threshold is taken as the orange dashed line in Fig.~1. We have also checked and confirmed that a more constraining threshold (red dashed line in Fig. 1) yields essentially identical results. 

For the structural measurements, we choose JWST wavebands $\lambda = 3.56\,\mu m$ (at $z=2-4$) and $\lambda = 2.00\,\mu m$ (at $z=0.5-2$); HST-CANDELS H-band: $\lambda = 1.65\,\mu m$ (at $z=0.5-2$); and SDSS: r-band $\lambda = 0.65\,\mu m$ (at $z=0.02-0.2$). This ensures that at all epochs probed, spanning 13\, Gyr of cosmic evolution, we probe approximately the red-optical to near-infrared in the rest-frame of galaxies. The exact wavebands used are, of course, limited by availability and the quality of structural measurements achieved. For example, in HST-CANDELS we use the reddest available band, to best approximate a mass distribution; but in the SDSS the $r$-band measurements are chosen for their superior quality. We find only a modest dependence on waveband for most ranges in JWST (see later in this section). Moreover, these small variations in rest-frame wavelength coverage could at most lead to more significant {\it differences} between the varying surveys (and our principal result is a lack of variation, see below).


\begin{figure*}
\begin{centering}
\includegraphics[width=0.8\textwidth]{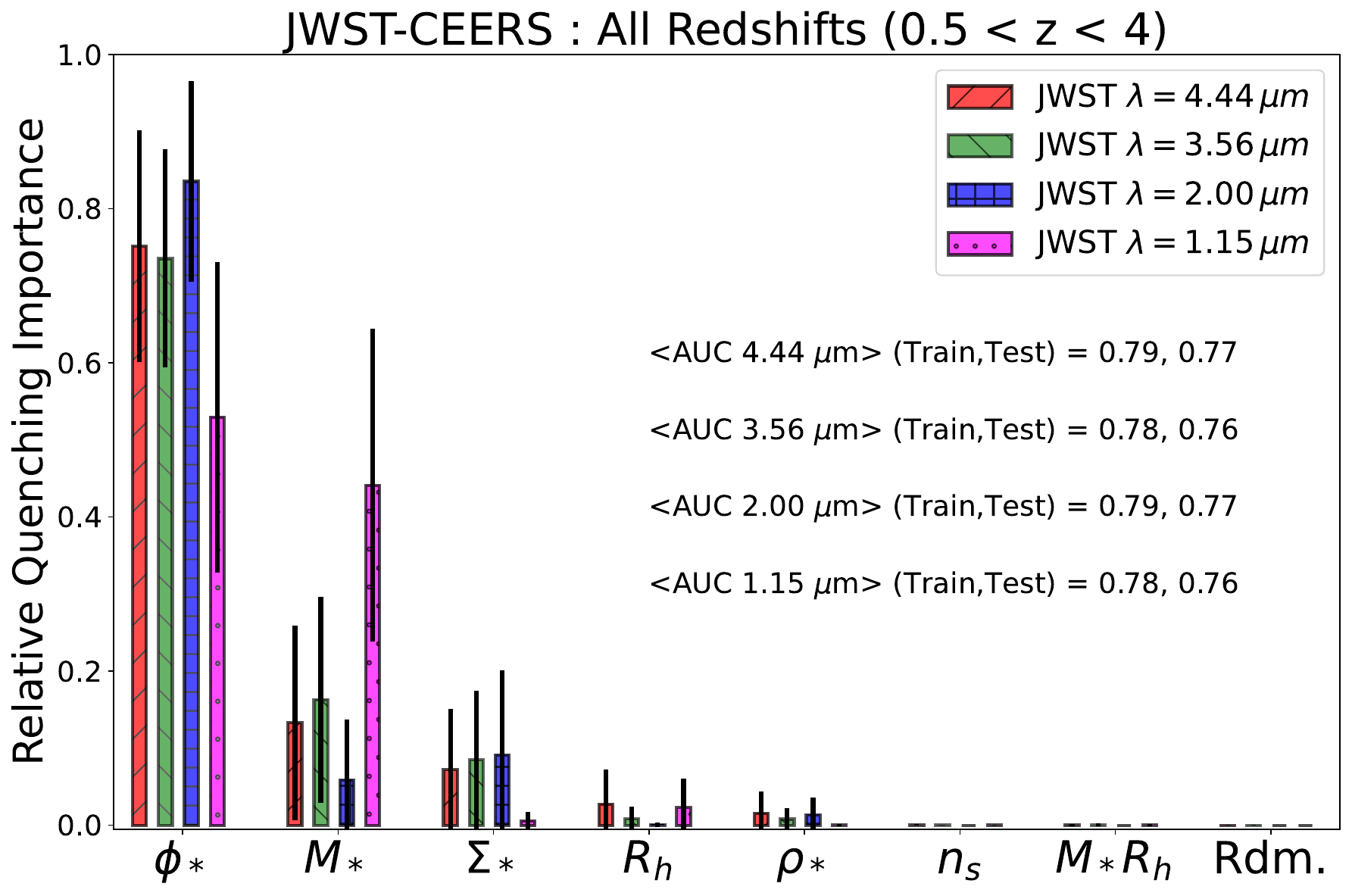}
\caption{Extended Random Forest classification of JWST-CEERS data. In this figure we increase the number of observable parameters used to train the classifier in Fig.~3, including Sersic index ($n_S$), surface density ($\Sigma_*$), and volume density ($\rho_*$). Additionally, we explore the impact of wavelength on the importance of the structural parameters to quenching (see legend). Here we analyse the entire redshift range in JWST-CEERS, since the importance of variables to quenching does not appear to change with cosmic time in Fig.~3. The stellar potential remains the clear best parameter for predicting quiescence in the larger parameter set, when measured in red bands, but becomes equivalent in importance to stellar mass in the bluest band available. This strongly implies that the value of $\phi_*$ is connected to the mass distribution (traced at longer wavelengths). When this measurement is heavily biased by the distribution of bright star forming regions (i.e. in the bluest band) it is less valuable for predicting quiescence. Note also that the uncertainties (measured from the variance of independent classification runs) increase with respect to Fig. 3. This is an expected consequence of adding more closely inter-correlated variables, which provides more pathways to segregation of galaxies into star forming and quiescent classes. See Table 3 in Appendix A for full details on the RF hyperparameter setup. }
\end{centering}
\end{figure*}

From Fig.~3 it is clear that the stellar gravitational potential, $\phi_*$, is by far the most predictive variable of galaxy quenching at all epochs. The stellar potential greatly outperforms stellar mass, galaxy size, and mass times radius (all of which perform similarly to a random variable). This is in remarkable agreement with the predictions from the EAGLE and TNG hydrodynamical simulations (see Fig.~2, and Bluck et al. 2023 for lower-$z$ simulations results). In the models this is a result of AGN feedback driving quenching, and central black holes growing in dense potential wells.

We emphasize the remarkable stability in the importance of parameters for predicting quenching in galaxies across 13\, Gyr of cosmic history seen in Fig.~3. This strongly suggests that quenching must be a stable mechanism operating across cosmic time. That is, at base, there is no fundamental observable difference between quenching at high or low redshifts. Again, this result is precisely predicted by hydrodynamical simulations (compare the results at $z<2$ in Bluck et al. 2023 to Fig.~3). This is a truly impressive achievement for the AGN feedback paradigm, which was designed and calibrated using only low redshift data.

Unlike in simulations, in observational data there are uncertainties on the measurements of parameters used to train the classifier. As such, it is important to carefully consider whether these effects could lead to an erroneous conclusion. The parameters used in Fig.~3 have been carefully chosen to mitigate this potential problem. 

The uncertainties on stellar mass are estimated to be $\sim$0.2 - 0.3\,dex, accounting for uncertainties in both SED fitting and the SED template libraries used (see Duncan et al. 2019). The uncertainties on galaxy sizes are $\sim$0.1 - 0.2\,dex (see Ormerod et al. 2023, in prep.). Since both $\phi_*$ and $M_* R_h$ are formed from linear combinations of mass and radius, their errors are given (by adding in quadrature) as: $\sim$0.25 - 0.35\,dex. In Bluck et al. (2022) we demonstrate that RF classification results are extremely stable for parameters with the same, or similar, relative errors, but that differential measurement uncertainty can lead to less well measured parameters performing worse than they should.

Crucially, in the example of Fig.~3, the worst measured parameters are $\phi_*$ and $M_* R_h$. This is a necessary consequence of these parameters taking the other two parameters as inputs\footnote{For a relationship $C = A/B$, the relative error on C is given by: $(eC/C) = \sqrt{(eA/A)^2 + (eB/B)^2)} > (eA/B) \,\, | \,\, (eB/B)$, where, e.g., $eA$ denotes the uncertainty on A (measured in units of variable $A$). Note also that it is the relative error which matters for the RF analysis, because we median subtract (and normalize by the inter-quartile range) all parameters to ensure a fair comparison (see Appendix A).}. Yet, the former clearly outperforms stellar mass and radius. Hence, this result cannot be explained by differential uncertainty. Furthermore, the relative uncertainties on $\phi_*$  and $M_* R_h$ are identical (each being the sum of the relative errors on $M_*$ and $R_h$ in quadrature). Yet, $\phi_*$ vastly outperforms $M_* R_h$. Therefore, these results cannot be explained, at any epoch, by measurement uncertainty. 

Furthermore,  the inclusion of $M_* R_h$ in the list of parameters clearly demonstrates that an arbitrary (non physically motivated) combination of mass and radius performs worse than mass and radius individually, and no better than a random variable. Yet, a carefully chosen theoretically motivated combination (i.e., $\phi_* \sim  M_{BH}$, according to simulations) performs as clearly the best predictor of quiescence in all galaxy surveys, and at all epochs. This motivates the, perhaps unusual, choice to include $M_* R_h$.

It is worth emphasizing this point a little further. An arbitrary combination of variables obscures the information contained within each component. For example, a phone number plus a random number is of no use for making contact. On the other hand, a well chosen combination can reveal great insight. For example, the square root of a phone number multiplied by itself does, of course, enable contact. In terms of the problem at hand, $\phi_*$ ($= M_* / R_h$) is of great predictive value for identifying quiescence, yet $M_* R_h$ is of no predictive power at all. Hence, one cannot claim that the superiority of $\phi_*$ in predicting quenching is a trivial consequence of it being based on $M_*$ and $R_h$. Rather, the superiority of $\phi_*$ is a consequence of it being a {\it valuable} combination of $M_*$ and $R_h$ (as predicted by simulations, see Section 4.1). Nonetheless, this does not necessarily imply that the combination is optimal. To test this further we next explore a greater number of variables.

In Fig.~4 we show an additional set of RF quenching classifications applied to the JWST-CEERS data. Here we expand further the list of parameters used to train the classifier, by including galaxy surface stellar mass density ($\Sigma_* = M_* / R_h^2$), galaxy volumetric mass density ($\rho_* = M_* / R_h^3$), and Sersic index ($n_S$). Additionally, we compare structural results in four wavebands, spanning the entire JWST-CEERS wavelength range. Note that in this figure we analyse the entire redshift range in JWST-CEERS together, since no significant differences as a function of redshift are seen in Fig. 3.

As in Fig.~4, $\phi_*$ is clearly found to be the best parameter for predicting quiescence in JWST-CEERS in the larger parameter space, and at all wavelength ranges. The only exception to this is in the bluest band, where $\phi_*$ and $M_*$ become indistinguishable. At the median redshift in JWST-CEERS, this band corresponds to rest-frame UV. As such it is no longer a good measurement of the underlying mass distribution in galaxies. Instead, it traces the location of bright, blue, massive young stars (which make up only a tiny fraction of the mass budget in essentially all massive galaxies).

As a result of the above, the reduction in importance of $\phi_*$ in the bluest JWST waveband is actually the exception which proves the rule. In the AGN feedback paradigm, the value of $\phi_*$ is that it measures the galaxy potential, which correlates strongly with black hole mass. However, once the half-light radius is no longer a good approximation of the half mass radius (as in the bluest bands which are contaminated by young stellar populations), this is no longer true, and the link to black hole mass is lost. Indeed, if the results of Fig.~4 were inverted, this would be exceptionally worrying, since it would suggest that the value in combining galaxy radius with stellar mass stems not from the importance of the gravitational potential but, rather, from a trivial tracing of young stars. Ultimately, the fact that this is emphatically not the case is very reassuring for the robust interpretation of these quenching results from JWST-CEERS.

In terms of interpretation, the results shown for simulations clearly show that a dominant dependence of quenching on $\phi_*$ is expected for AGN quenching, which ultimately depends on $M_{BH}$. There are no dynamical black hole mass estimates for our high-$z$ samples. Hence, the best that can currently be done to test the AGN feedback paradigm at these epochs is to see how quenching depends on measurable photometric parameters. This provides clear indirect support for the AGN feedback paradigm (the phenomenology of Figs. 2 and 3 are identical). Additionally, note again that a dependence of quenching on accretion rate (and hence AGN luminosity and detection) is not predicted to be evident in models. Therefore, the current approach is essentially the optimal methodology for testing the AGN feedback paradigm at high redshifts.


\begin{figure*}
\begin{centering}
\includegraphics[width=0.95\textwidth]{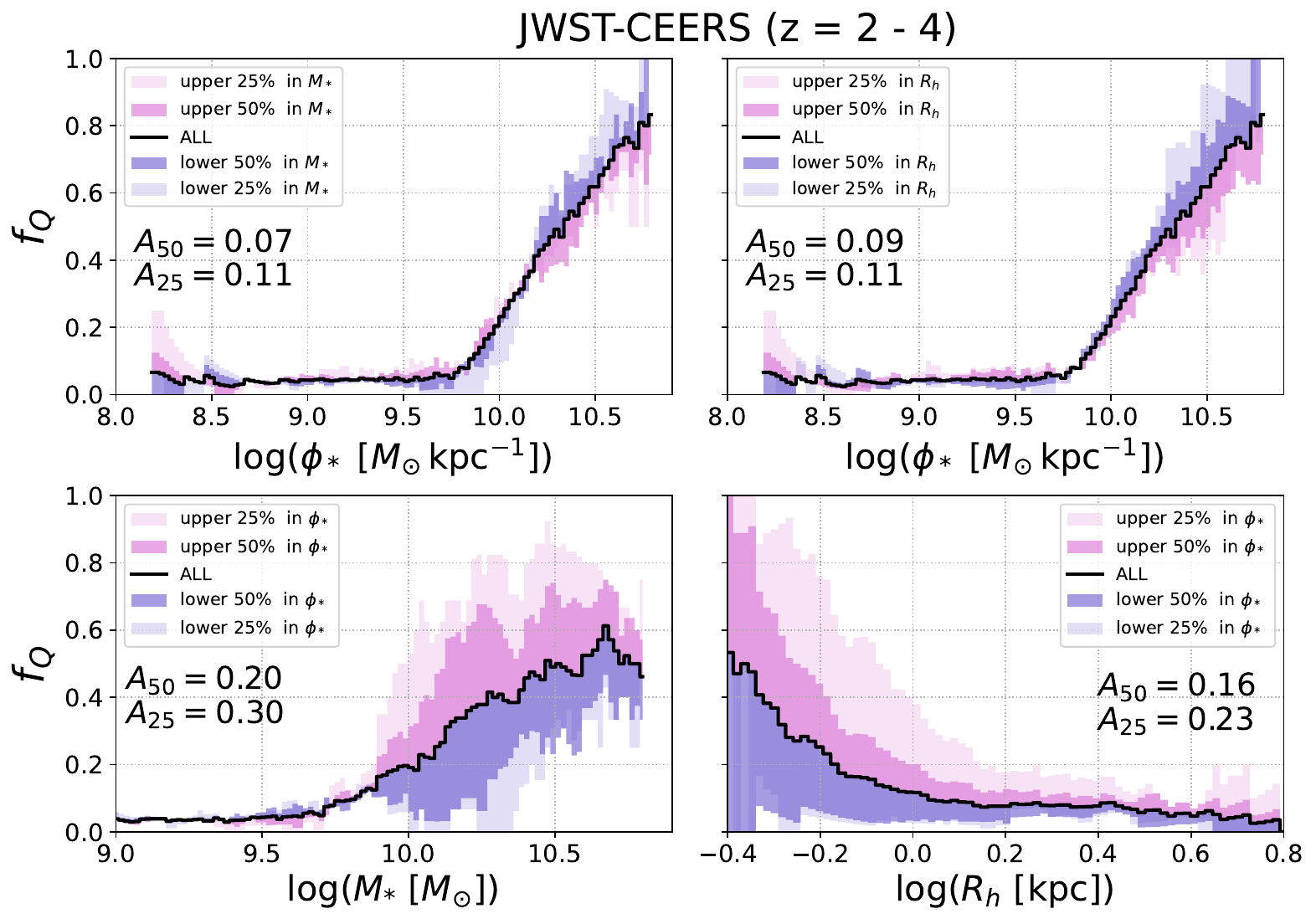}
\caption{Area statistics analysis to check the validity of the RF results in JWST-CEERS at $z = 2 - 4$. The quenched fraction relationships are displayed as a function of stellar potential (top panels), stellar mass (lower-left panel), and galaxy radius (lower-right panel). Each quenched fraction relationship is subdivided into percentile ranges of a secondary parameter (as labelled by the legends). For example, the top-left panel displays the quenched fraction relationship with $\phi_*$ split into percentile ranges of $M_*$ at each step. The bottom-left panel inverts this, showing the quenched fraction relationship with $M_*$ split into percentile ranges of $\phi_*$ at each step. It is clear that varying $\phi_*$ at fixed $M_*$ or $R_h$ engenders far more variation in the quenched fraction than the other way around. This strongly indicates that $\phi_*$ is a more fundamental parameter for quenching (exactly as concluded in the RF classification analyses, see Figs. 3 \& 4). We quantify this effect with the area statistics ($A_{50}$ and $A_{25}$, which show the area subtended between the upper and lower 50\% and 25\% of the data at each bin, respectively, and are displayed on each panel). Higher values of these statistics indicate less tight relationships.}
\end{centering}
\end{figure*}

\subsubsection{Area Statistics Check}

\noindent The results from the RF classification analyses of JWST-CEERS galaxies (shown above) are extremely clear. The stellar potential, $\phi_*$, is identified as the most important observable variable for predicting quiescence in galaxies at all epochs studied, spanning $\sim$13\,Gyr of cosmic history. These results are precisely as predicted in the AGN feedback driven quenching paradigm for these photometric parameters (see Section 4.1, and Piotrowska et al. 2022; Bluck et al. 2023). However, given both the novelty and importance of the results of the previous sub-section, it is sensible to check the validity of the RF classification conclusions via a simpler, and arguably more transparent, technique.

In Fig.~5 we present the quenched fraction relationships with $\phi_*$ (top panels), $M_*$ (lower-left panel), and galaxy half-light radius (lower-right panel). Results are shown for the highest JWST-CEERS redshift range used in the RF classification above ($2 < z < 4$). Each quenched fraction relationship is subdivided into separate relationships for percentile ranges of a third variable. For example, in the top-left panel the $f_Q - \phi_*$ relationship is displayed in percentile ranges of $M_*$ (i.e., upper 25\%, upper 50\%, lower 50\%, lower 25\%). The lower-left panel inverts this, displaying the $f_Q - M_*$ relationship in ranges of $\phi_*$. Clearly, varying $\phi_*$ at fixed $M_*$ engenders a far greater impact on the quenched fraction than the other way around, implying that the stellar potential must be more fundamentally linked to quiescence than stellar mass.

We quantify the tightness of each relationship using the area statistics approach (see Bluck et al. 2016, 2020a, 2022). We integrate the quenched fraction from the upper to the lower percentile range (for upper and lower 50th and 25th percentiles in each secondary variable, labelled by the legend in each panel). The integration area is then normalized by the total area displayed in each panel, for a fair comparison. This method yields a relative tightness, dependent upon a third variable (since an absolute tightness is not obtainable for a simple fraction). This quantifies, e.g., the impact of varying $M_*$ at fixed $\phi_*$ (top-left panel), and $\phi_*$ at fixed $M_*$ (lower-left panel). In the right panels, the same process is applied to $R_h$ and $\phi_*$. As such, Fig. 5 should be `read' column-wise to compare the relative impact of each pair of variables to quenching. 

We estimate statistical uncertainties on the area statistics through bootsrapped random sampling of the data. Typical uncertainties are found to be $\sim$0.02. However, the value of the area statistics vary systematically with binsize, which we hold constant in all plots to remove this issue. Additionally, we have checked that the general trends (i.e. which parameter is tightest) is invariant to reasonable binsize choices. Consequently, we conclude that the improvement in parameterization of quenching with $\phi_*$ over both $M_*$ and $R_h$ is highly significant, despite the increase in relative error on the former with respect to the latter (see above), and is analysis methodology independent.

From Fig.~5, it is clear that the $f_Q - \phi_*$ relationship is much less impacted by either $M_*$ or $R_h$, than the other way around. This clearly indicates that $\phi_*$ is the most fundamental (i.e. least impacted by other variables) parameter for driving quiescence in massive galaxies in JWST-CEERS data at very high redshifts, out of the list considered. Hence, the conclusions from the area statistics approach exactly matches the conclusions found in the RF classification analyses above. Additionally, the area statistics approach yields consistent results with the RF classifications at lower redshifts as well (see Bluck et al. 2023).

At this point one might wonder what the need of the RF classification is. The answer to this is twofold. First, the RF classifier can simultaneously control for an arbitrary number of parameters, whereas the area statistics approach can consider only two at a time. Second, the classification analysis explicitly solves the problem of identifying quiescent objects, whereas the area statistics approach only suggests an approach for doing so. In summary, the area statistics offer a simple, but useful, check on the results of the more sophisticated RF classification, but are not a substitute for it. \\

\subsubsection{Quenching in the $\phi_* - M_*$ Plane}

\noindent In Fig. 6 we present the location of star forming and quiescent galaxies from JWST-CEERS in the $\phi_* - M_*$ plane, shown separately for intermediate redshifts ($0.5 < z < 2$), high redshifts ($2 < z < 4$), and ultra-high redshifts ($4 < z < 8$). Note that this is the first time we introduce an analysis of ultra-high redshift systems. The reason for this is that the number of quiescent galaxies is far too low in JWST-CEERS at these epochs for the previous statistical analyses (i.e., RF classification in Section 4.2.1; and area statistics in Section 4.2.2). As such, our conclusions for the ultra-high redshift galaxies will be preliminary.

In all panels of Fig. 6 the location of star forming galaxies (identified as objects above the dashed orange line in Fig. 1) are shown by blue density contours. The median relationship for star formers is shown by a solid blue line. Clearly, at all epochs there is a very strong relationship between stellar mass and stellar potential, such that more massive galaxies have deeper potential wells (with correlations, $\rho = 0.77, 0.78, 0.91$ for low, high, and ultra-high redshifts, respectively). However, crucially for this analysis, there is physically meaningful scatter in the relationship owing to variation in galaxy size (and hence mass distribution) at a fixed stellar mass.

We also show the location of quenched (below the orange line in Fig. 1) and fully-quenched (below the red line in Fig. 1) high and low mass galaxies on the $\phi_* - M_*$ plane in Fig. 6 (color coded as described in the legends). Note that we include galaxies in the high mass sample if $>$30\% of their PDF extends into the high mass regime. To guide the eye, we overlay dashed black lines to indicate separation into high and low stellar mass systems (located at $M_* = 10^{10} M_\odot$) and show a separation into deep and shallow potentials (located at $\phi_* = 10^{10} M_{\odot}/{\rm kpc}$). This separates the parameter space into four quadrants of interest for further discussion.

At low masses, quiescent objects trace the median star forming relationship at all epochs. This suggests that there is no impact of varying $\phi_*$ at a fixed $M_*$ for these systems. Utilizing the interpretation of simulations, this implies that low mass quenching is unrelated to AGN feedback (as expected). Consequently, these quiescent objects are most likely the result of environmental quenching in the early Universe.

For high mass quenched galaxies in Fig. 6, we see an offset in $\phi_*$ at fixed $M_*$, such that quiescent galaxies reside in deeper potential wells for their stellar masses than star forming systems. This offset can be naturally explained in the AGN feedback paradigm by these systems hosting more massive central black holes, which will have injected more energy into their circum-galactic media. Interestingly, we see evidence of this offset all the way out to the most extreme redshift range ($z = 4 - 8$).


\begin{figure*}
\begin{centering}
\includegraphics[width=0.54\textwidth]{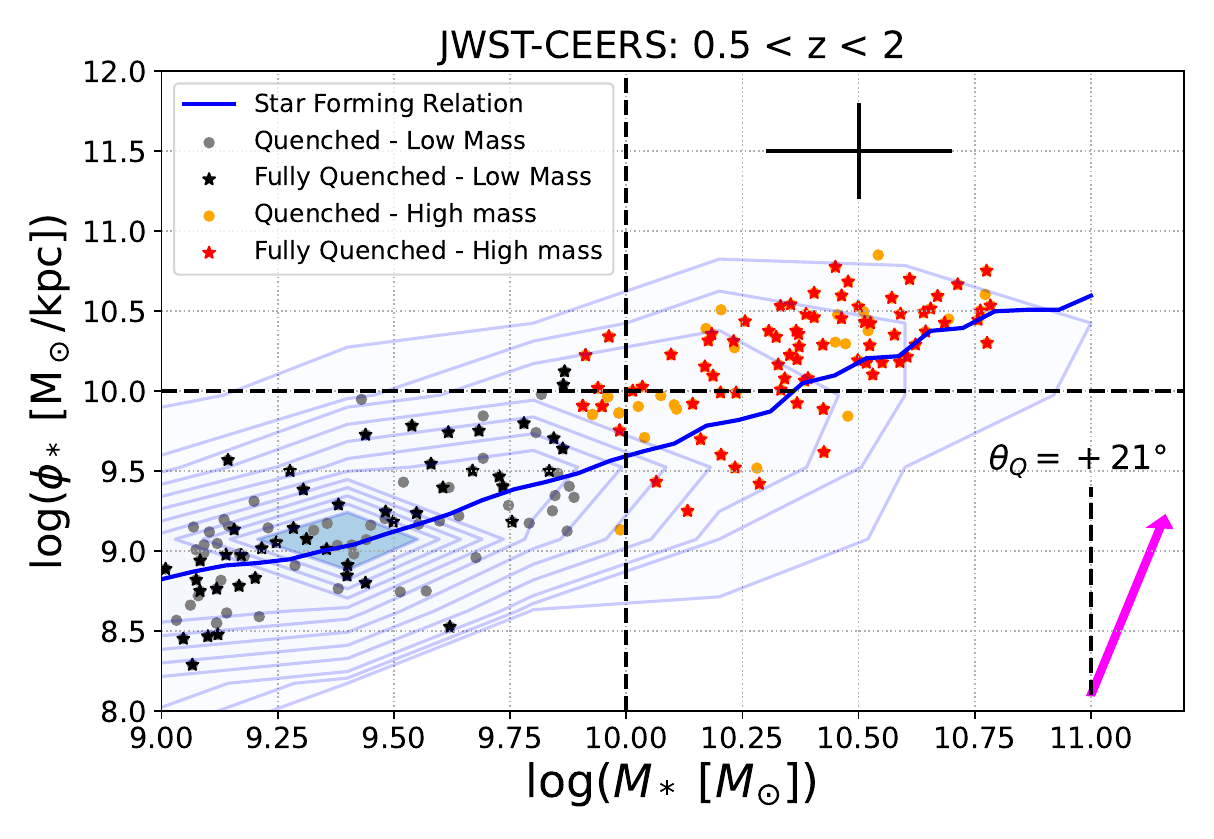}
\includegraphics[width=0.54\textwidth]{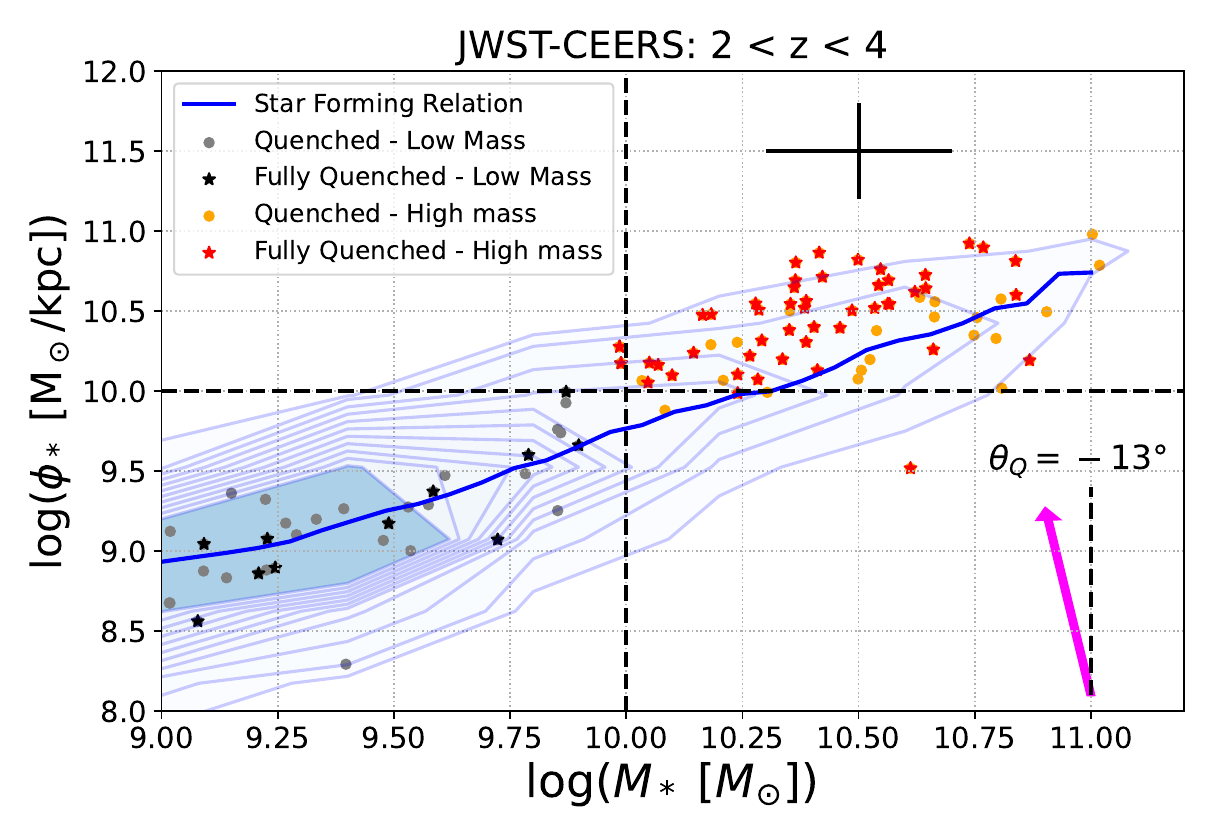}
\includegraphics[width=0.54\textwidth]{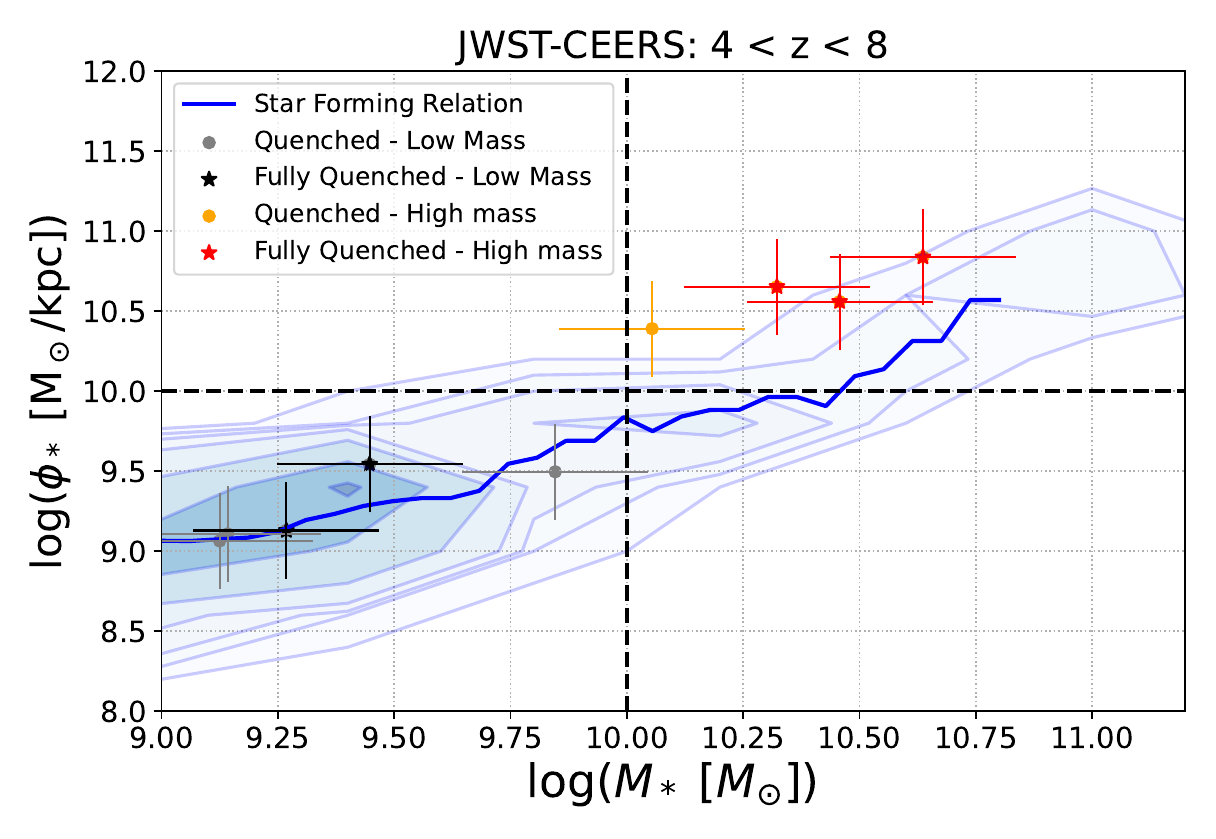}
\caption{The stellar potential - stellar mass ($\phi_* - M_*$) relationship in JWST-CEERS. Results are shown separately for intermediate redshifts (top panel), high redshifts (middle panel), and ultra-high redshifts (bottom panel). On each panel the location of star forming galaxies (above the orange dashed line in Fig. 1) are shown by blue density contours. The median $\phi_* - M_*$ relationship for star formers is indicated by a solid blue line. There is clearly a very strong relationship at all epochs, whereby more massive galaxies have deeper gravitational potentials, with physically meaningful scatter engendered by varying galaxy size at a fixed mass. The location of high mass fully-quenched (red), high mass quenched (orange), low mass fully-quenched (black), and low mass quenched (grey) galaxies are shown by data points. The typical uncertainty on these measurements is indicated by the black error bars on the top two panels, and by individual error bars in the lower panel. At all epochs, high mass quiescent systems are offset to higher $\phi_*$ values at fixed $M_*$; whereas low mass quiescent galaxies are not significantly offset in this diagnostic relative to star formers. For intermediate and high redshifts, we quantify the effect for high mass systems with the quenching arrow (see text), which shows the optimal direction to move in a 2D plot in order to quench. Since $\|\theta_Q\| < 45^\circ$, we conclude that $\phi_*$ is more important to quenching than $M_*$ (as in the RF and area statistics analyses above). }
\end{centering}
\end{figure*}


\begin{figure*}
\begin{centering}
\includegraphics[width=0.49\textwidth]{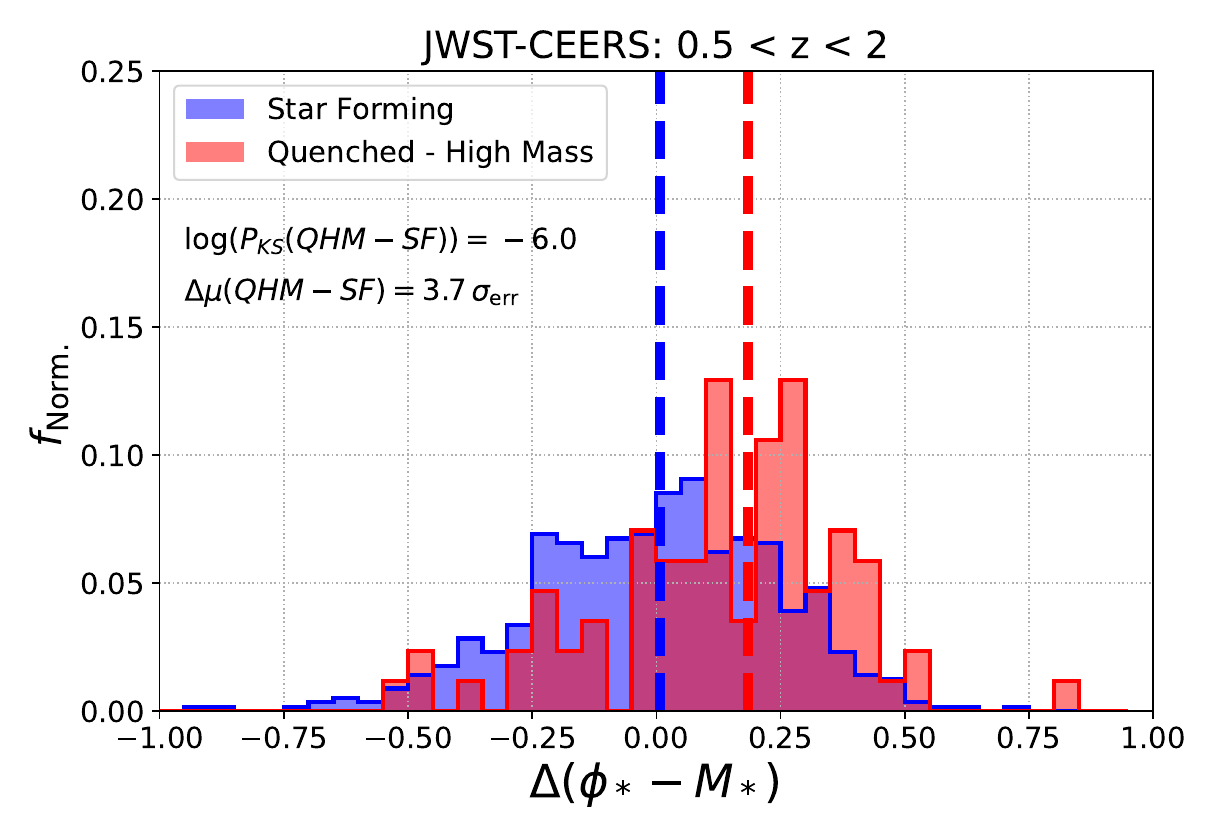}
\includegraphics[width=0.49\textwidth]{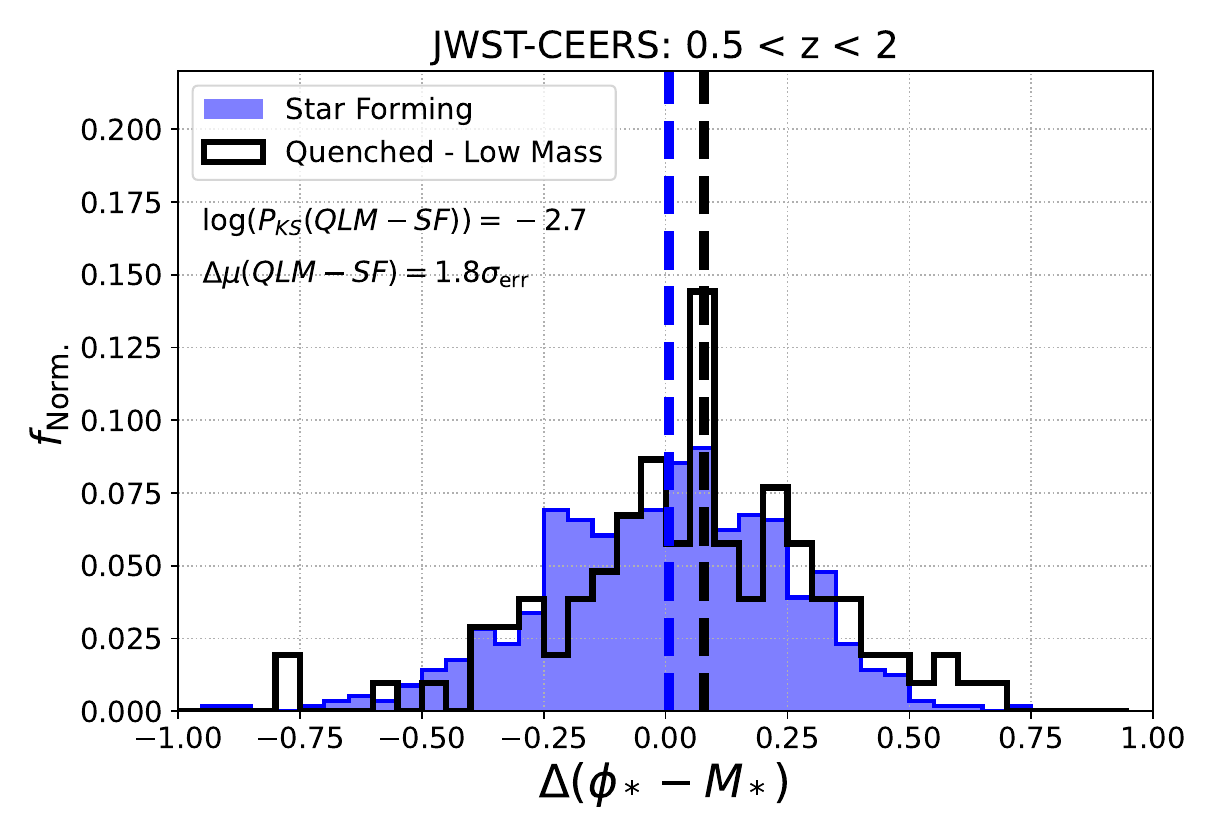}
\includegraphics[width=0.49\textwidth]{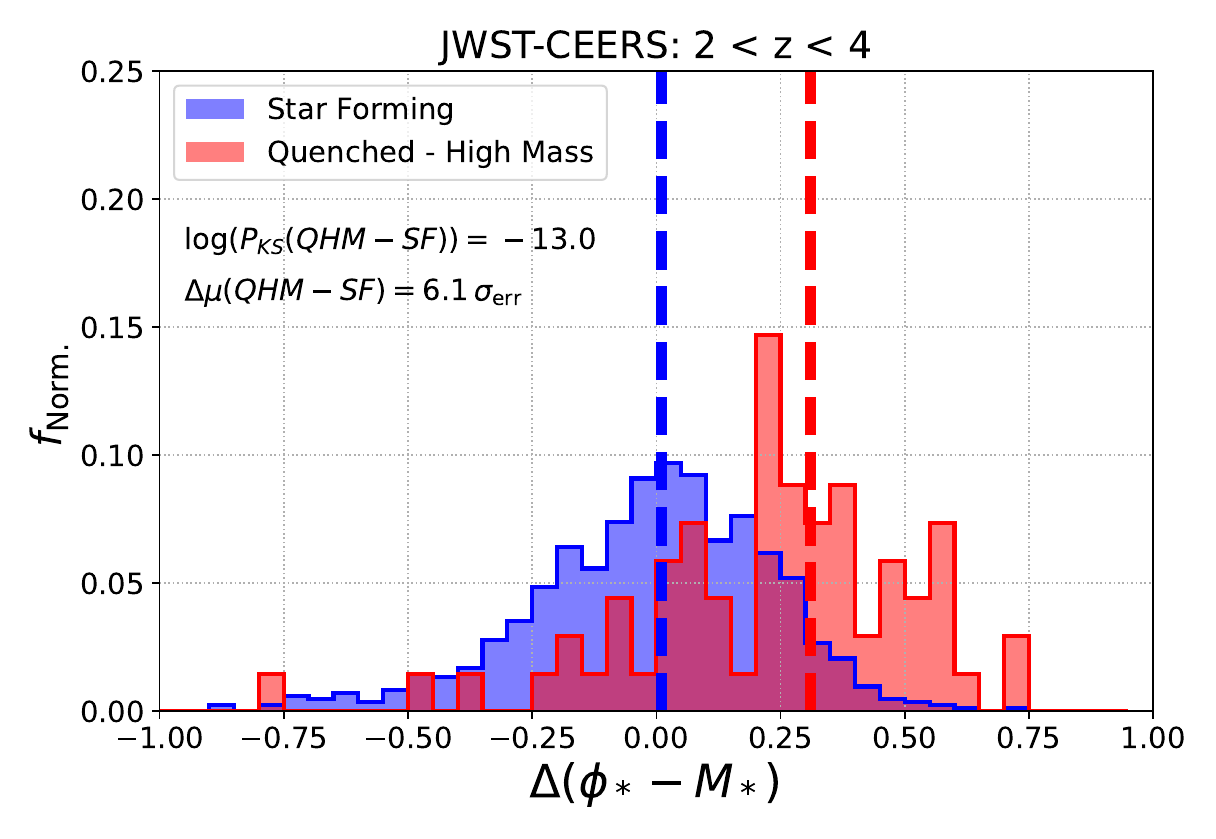}
\includegraphics[width=0.49\textwidth]{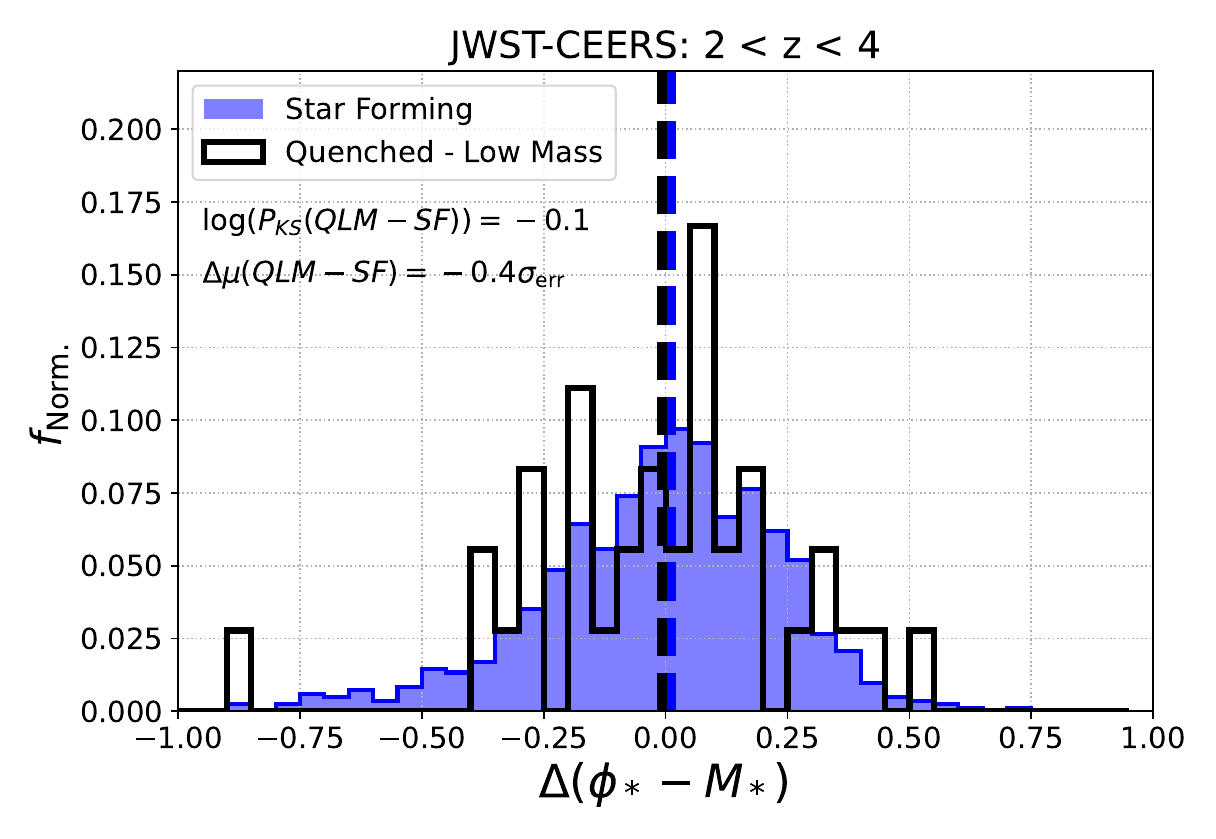}
\includegraphics[width=0.49\textwidth]{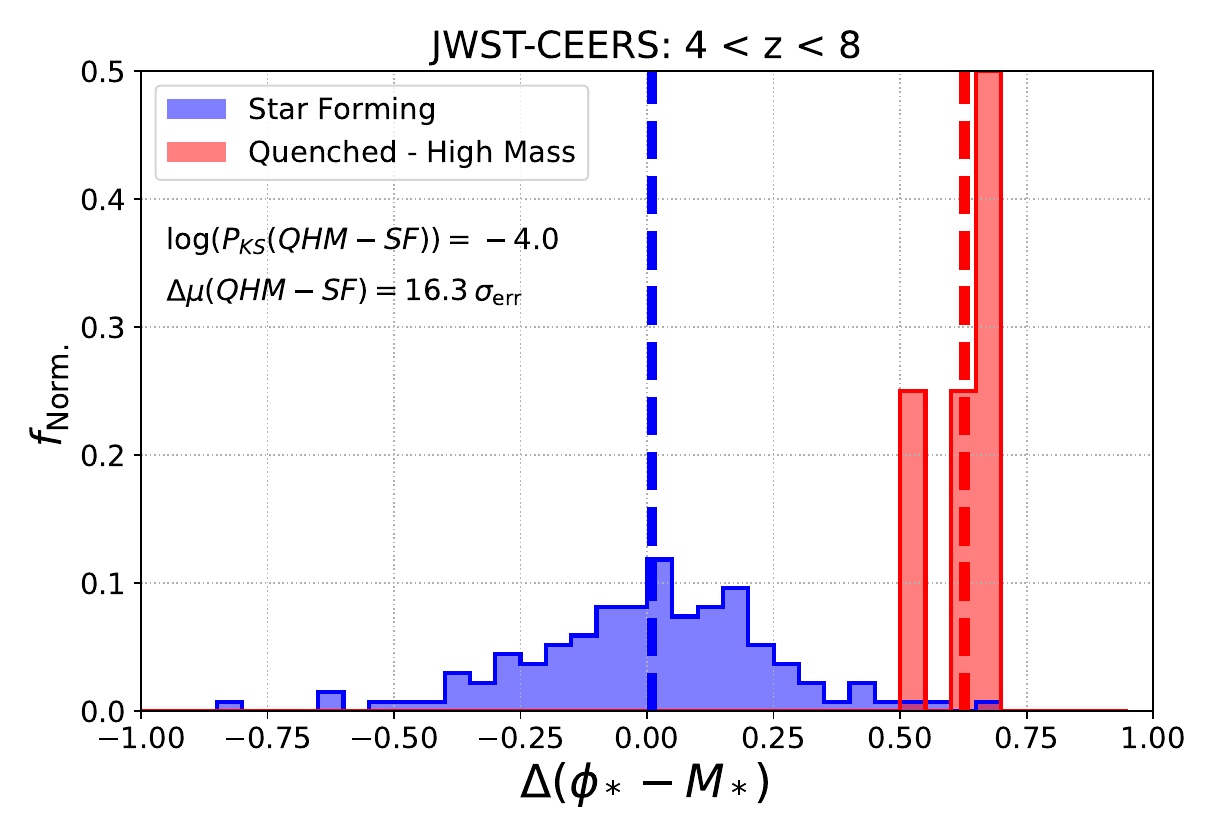}
\includegraphics[width=0.49\textwidth]{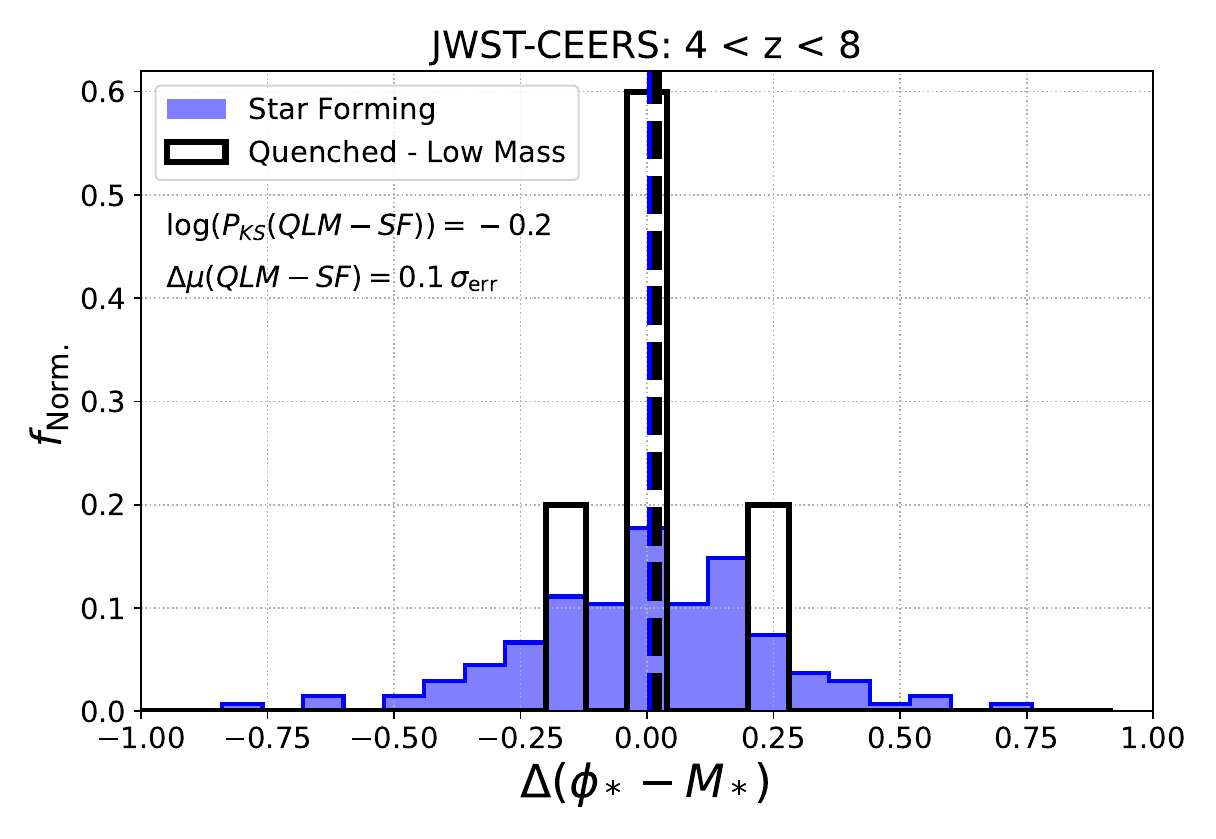}
\caption{Distributions in offsets from the star forming $\phi_* - M_*$ relationship ($\Delta(\phi_* - M_*)$). Offsets are measured relative to the star forming median relationship at each epoch (blue solid lines in Fig. 6). The left panels show a comparison of quenched galaxies to star forming galaxies at high mass ($M_* > 10^{10} M_\odot$), with right panels showing the same comparison for low mass galaxies ($M_* < 10^{10} M_\odot$). The top row displays results for intermediate redshifts, the middle row for high redshifts, and the bottom row for ultra-high redshifts. It is clear that quiescent galaxies are offset to higher $\phi_*$ at high masses, relative to star formers at a fixed stellar mass. Additionally, it is also clear that there is no significant offset for low mass systems. We quantify these effects with two statistics (displayed on the individual plots). First, we quantify the significance of the difference between each pair of distributions by the Kolmagorov-Smirnov $p$-value ($P_{KS}$). Second, we display the difference in mean values of each pair of distributions, in units of the combined error on the mean ($\Delta \mu \, [\sigma_{\rm err}]$). We note a trend whereby high mass quiescent galaxies become more offset to higher $\phi_*$ at higher redshifts (relative to star formers at a fixed stellar mass). }
\end{centering}
\end{figure*}

As yet another test on the relative importance of stellar mass and stellar potential for quenching high mass galaxies, in Fig. 6 we plot the quenching angle ($\theta_{Q}$). This is defined as the arctangent of the ratio of partial correlations, i.e. treating each partial corelation as a basis vector. Explicitly, we compute:

\begin{equation}
\theta_Q \equiv -\tan^{-1}\bigg\{ \frac{\rho (\mathrm{sSFR}/(1/t_H(z)) - \phi_* | M_*)}{\rho (\mathrm{sSFR}/(1/t_H(z)) - M_* | \phi_*)} \bigg\}
\end{equation}

\noindent where the numerator indicates the rank-ordered partial correlation between normalised sSFR and stellar potential, at fixed stellar mass; and the denomenator indicates the rank-ordered partial correlation between normalised sSFR and stellar mass, at fixed stellar potential. The sSFR normalization is used to account for redshift evolution in the quenching threshold (see Fig. 1), and the negative sign ensures that the angle points in the direction of increasing quiescence (rather than star formation). The partial correlations are given as standard by:

\begin{equation}
\rho_{A-B | C} = \frac{\rho_{AB} - \rho_{AC} \cdot \rho_{BC}}{\sqrt{1-\rho_{AC}^2}\sqrt{1-\rho_{BC}^2}}
\end{equation}

\noindent for three arbitrary parameters, $A, B,$ and $C$. For example, $\rho_{AB}$ indicates the Spearman rank correlation between variables $A$ and $B$, and $\rho_{A-B | C}$ indicates the partial correlation between $A$ and $B$, controlling for $C$. See Bluck et al. (2020a) for a full explanation of the quenching angle, and applications at low-$z$. Ultimately, this statistic quantifies the optimal route through a 2D parameter space to lower sSFR.

A quenching angle of 0$^\circ$ indicates pure $\phi_*$ dependence, an angle of 90$^\circ$  indicates full $M_*$ dependence, with 45$^\circ$  indicating an equal contribution of both. Since $\|\theta_Q\| < 45^\circ$ in both panels where we have the statistics to run this test, we conclude that $\phi_*$ is more important to quenching than $M_*$ in JWST-CEERS (as concluded in the RF and area statistics analyses above).

\subsubsection{Distributions in offsets from the $\phi_* - M_*$ relation}

\noindent In Fig. 7 we present the distributions in the offsets of quiescent galaxies from the $\phi_* - M_*$ median relation (defined for star forming objects, shown as blue solid lines in Fig. 6). Left panels show results for high mass quiescent galaxies, with right panels showing results for low mass quiescent galaxies. It is clear that high mass quenched objects are offset to higher $\phi_*$ at fixed $M_*$, relative to star forming galaxies. Conversely, there is no apparent offset in the low mass quenched systems, relative to star forming galaxies.

We quantify the significance and magnitude of the offsets (or lack thereof) via two statistics displayed on each panel of Fig. 7: (i) the probability of being drawn from the same distribution via the Kolmagorov-Smirnov (KS) test ($P_{KS}$); and (ii) the difference in mean offset between star forming and quenched populations, expressed in units of the combined error on the means of the two distributions ($\Delta \mu \,\, [\sigma_{\rm err}]$).


\begin{figure}
\begin{centering}
\includegraphics[width=0.49\textwidth]{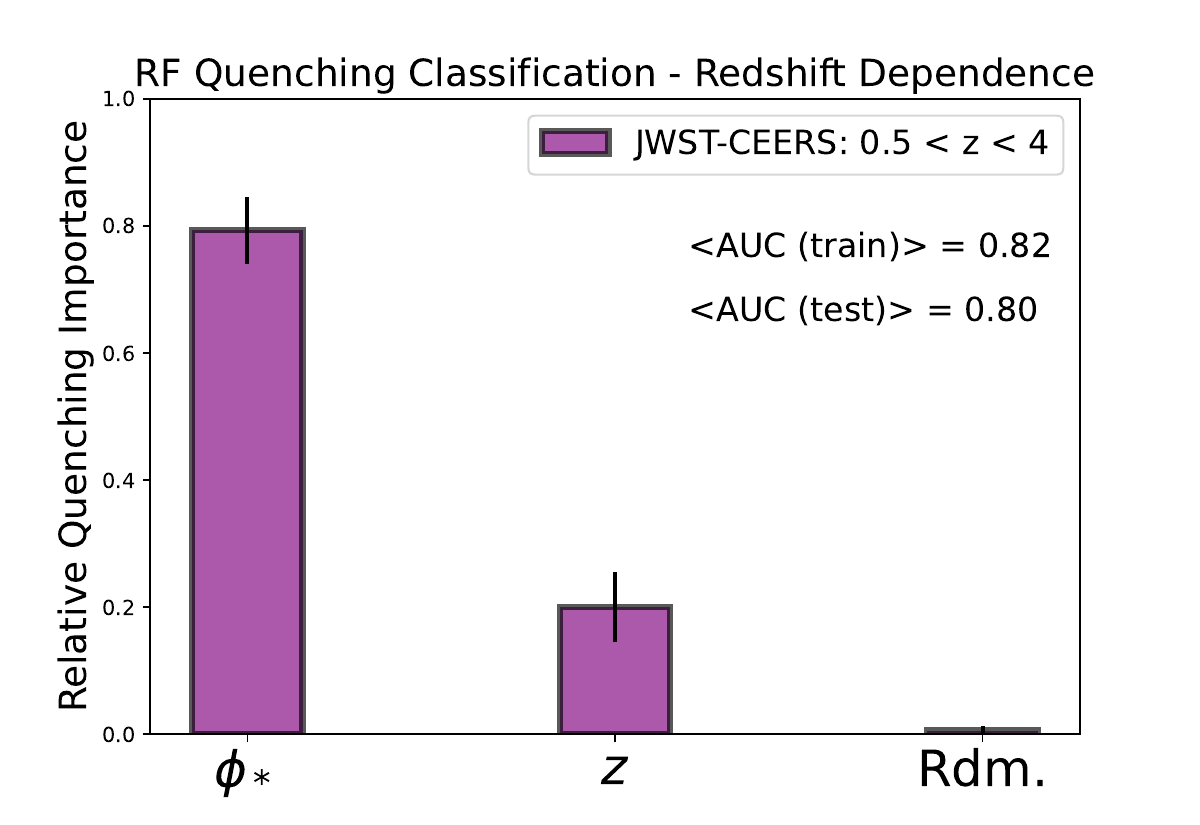}
\caption{RF classification analysis to predict the existence of quenched galaxies in the full JWST-CEERS redshift range, utilising only $\phi_*$, $z$, and a random variable. It is clear in this analysis that high mass galaxy quenching has a significant secondary dependence on redshift, even after controlling for the dominant parameter ($\phi_*$). This suggests evolution in the $f_Q - \phi_*$ relationship (see Fig. 9).}
\end{centering}
\end{figure}


\begin{figure*}
\begin{centering}
\includegraphics[width=0.95\textwidth]{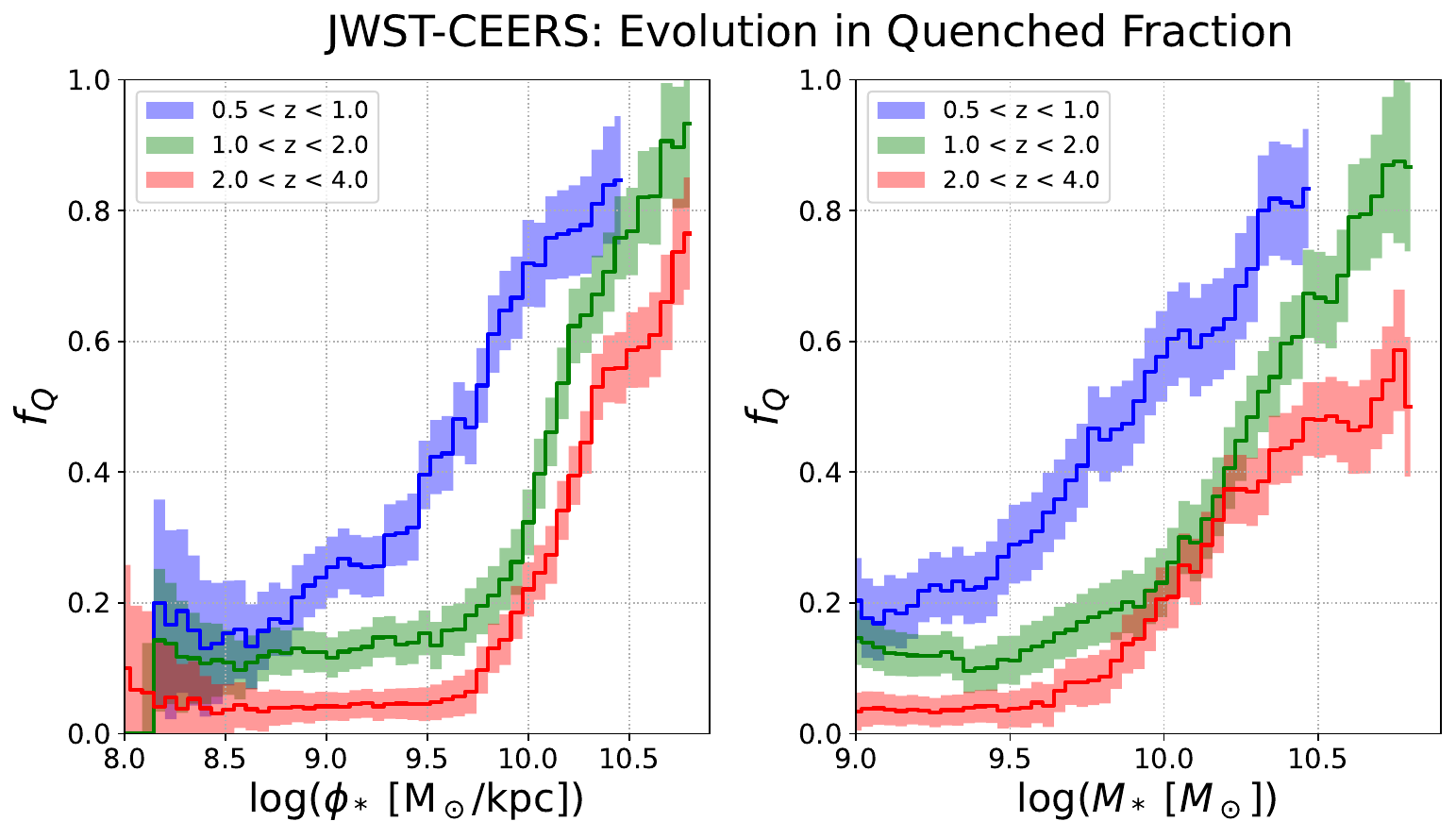}
\caption{Evolution in the quenched fraction relationship with stellar potential (left panel) and stellar mass (right panel). Solid lines show the quenched fraction as a function of $\phi_*$ and $M_*$, with shaded regions showing the 1$\sigma$ uncertainty in the quenched fraction (colored by epoch as displayed in the legends). Uncertainties are inferred from Poisson counting statistics. It is clear that there is a strong positive trend of quenched fraction with both stellar mass and stellar potential, which evolves with redshift such that higher values of both $\phi_*$ and $M_*$ are required at earlier epochs in order to quench. }
\end{centering}
\end{figure*}

High mass galaxies have significant offsets from the star forming $\phi_* - M_*$ relationship with confidence (probability of being spurious) ranging from 1/5000 (ultra-high redshifts) to 1/$10^{13}$ (high redshifts), with intermediate redshift galaxies being offset to a confidence of one in a million. Interestingly, the largest magnitude in offset is seen at the highest redshifts, and systematically declines to lower redshifts. This hints at a possible trend whereby in order to transition to quiescence in the very early Universe, galaxies must have preferentially deeper potential wells for their stellar masses, compared to later epochs. Within the AGN feedback paradigm, this would be interpreted as these systems forming exceptionally massive central black holes for their stellar masses, leading to enhanced AGN feedback integrated over their histories (see Section 5.1). However, better statistics at the highest redshifts are needed to confirm these interesting preliminary results.

On the other hand, for low mass quiescent galaxies, no significant offsets from the star forming $\phi_* - M_*$ relationship are seen at any epoch. This strongly suggests that this low mass quenching is unrelated to both stellar mass and gravitational potential, and hence also likely unrelated to black hole mass and AGN feedback. These quiescent systems are almost certainly a result of environmental, rather than intrinsic, quenching mechanisms (see, e.g., Bluck et al. 2016, 2020b; Goubert et al. in prep. for low-$z$ analogues). Hence, we find evidence for both high mass intrinsic quenching and low mass environmental quenching in JWST out to the most extreme redshifts ever probed. However, note that our results are preliminary in the sense that we have a photometric only sample with limited parameter availability, and have a modest sample size at the highest redshifts. Both of these limitations will be improved with upcoming JWST studies in the coming few years.

\subsubsection{Evolution in the $f_Q - \phi_*$ Relation}

\noindent In Section 4.2.1 we find that the best parameter for predicting quenching is stable throughout cosmic time (see back to Fig. 3). However, this does not mean that there is no explicit dependence of quenching on redshift, only that any such evolution must leave $\phi_*$ as the most predictive parameter at any given epoch. To explore the possibility of redshift dependence on quenching directly, in Fig. 8 we perform an RF classification analysis for the full JWST-CEERS range, comparing the dominant quenching parameter at each epoch ($\phi_*$) to redshift and a random number. It is clear that whilst $\phi_*$ remains the dominant parameter, there is a significant secondary dependence on redshift in the multi-epoch sample (which is much superior to random). This implies the need for evolution in the $f_Q - \phi_*$ relationship with cosmic time.

We test this expectation explicitly in Fig. 9 (left panel), where we show the $f_Q - \phi_*$ relationship for JWST-CEERS galaxies in bins of redshift (as labelled by the legends). As predicted by the RF analysis, there is significant evolution in $f_Q - \phi_*$ relationship, such that higher $\phi_*$ values are required to reach the same level of quenching at earlier cosmic times. For comparison, we also show the redshift evolution in the (less fundamental) $f_Q - M_*$ relation. Again, this shows marked evolution, such that higher masses are needed to quench galaxies at earlier epochs.

These results can also be explained elegantly within the AGN feedback paradigm. In simple models of structure formation, the accretion rate into dark matter haloes scales with the physical density of baryons in the intergalactic medium, i.e. proportional to $(1+z)^3$. Hence, at earlier cosmic times there is greater accretion into dark matter haloes, at a fixed total mass. Thus, in order to shut off this accretion (and prevent the hot gas atmosphere from collapse), greater energetic feedback from AGN will be required, leading to higher black hole mass quenching thresholds (and hence higher $\phi_*$ values) at earlier epochs. In turn, the $\phi_* - M_*$ relationship (Fig. 6) requires that there will be redshift dependence on the $f_{Q} - M_*$ relation as well, which is a non-causal reflection of the underlying evolution in the black hole quenching threshold. See Bluck et al. (2022) for an analytic derivation of this redshift dependence in a semi-analytic model.

\section{Discussion}

\subsection{AGN Quenching in the Early Universe}

\noindent A comparison of the results from Section 4.2 (observations) with Section 4.1 (simulations) clearly demonstrates that AGN feedback {\it can} explain the statistical properties of galaxy quenching across cosmic time. Indeed, the direct predictions from EAGLE and TNG are met precisely in observations across 13\,Gyr of cosmic history, incorporating data from three major galaxy surveys: (i) JWST-CEERS; (ii) HST-CANDELS; and (iii) the SDSS.

In both EAGLE and TNG, black hole mass is predicted to act as the most important variable for quenching at very early cosmic times, and remain stable throughout cosmic history to the present epoch (see Fig.~2; and Bluck et al. 2023; Piotrowska et al. 2022). Crucially, black hole accretion rate (and hence AGN luminosity and detection) are not predicted to be strongly connected to quiescence at any epoch. This is in spite of the fact that AGN feedback is explicitly the cause of high mass quenching in both simulations (see Schaye et al. 2015; Crain et al. 2015; Weinberger et al. 2017, 2018; Zinger et al. 2020). This implies that searching for the location of AGN on the star forming main sequence is not a productive route to testing the AGN feedback paradigm, despite its ubiquity in the observational literature (see Piotrowska et al. 2022; Ward et al. 2022; Bluck et al. 2023 for further discussion).

The reason for this is that in order for a galaxy to remain quiescent for long periods of time, cold gas inflow into the system must be prevented. As such, it is not enough to strip a galaxy of its interstellar medium (ISM). Hence, powerful energetic feedback (often referred to as the `quasar mode'; c.f. Hopkins et al. 2006; 2008) is not sufficient to ensure a fully quenched system. Additionally, even without ISM removal, a galaxy can still quench if gas inflows are permanently switched off. In this scenario, galaxies will continue to transform their ISM into stars, and quench only once it is fully depleted. This latter (slow) quenching leads to the prediction that the stellar metallicities ought to rise during transition to quiescence (which is confirmed in Peng et al. 2015; Trussler et al. 2020; Bluck et al. 2020b).

Therefore, high Eddington ratio, energetic AGN (which are by far the easiest type of AGN to identify) are neither necessary nor sufficient to quench massive galaxies. The essential condition of massive galaxy quenching in modern simulations is the formation of low Eddington ratio AGN which act over cosmological times, preventing cooling of the CGM and, hence, gas accretion into the system. These conditions occur when black hole masses become very large, and the predominant accretion mechanism switches from cold- to hot-mode (e.g., Sijacki et al. 2007; Fabian 2012; Weinberger 2018; Terrazas et al. 2020; Zinger et al. 2020; Piotrowska et al. 2022). Observationally, this mode of AGN are extremely hard to detect. However, their long term effects on the hot gas halo are clearly seen in galaxy clusters (see Hlavacek-Llarondo et al. 2015, 2017, 2018; Brownson et al. 2019; Jones et al. 2023). Additionally, low luminosity core radio detection is identified preferentially in high mass quenched galaxies in the local Universe (see, e.g., Heckman \& Best 2014; Hickox et al. 2009), which is obviously consistent with this paradigm.

The total energy released into galactic haloes through long-term, low luminosity AGN heating in the `preventative-mode' is directly proportional to the mass of supermassive black holes (see Bluck et al. 2020a for a modern derivation; and Soltan et al. 1982; Silk \& Rees 1998 for the original arguments). Moreover, it is completely uncorrelated with the current AGN accretion rate, AGN bolometric luminosity and, hence, AGN detection in any given waveband. Ultimately, this reasoning explains the phenomenological results from simulations in Section 4.1, and in Bluck et al. (2023); Piotrowska et al. (2022); and Ward et al. (2022). Contemporary AGN feedback models explicitly predict that quiescence ought to intrinsically depend on black hole mass, not accretion rate or luminosity. As such, the correct method to test the AGN feedback paradigm is {\it not} to look for the location of AGN on the star forming main sequence (as many prior observational works have done), but rather to look for a connection between quiescence and the fossil record of past accretion, i.e. black hole mass (or its best available proxies).

In Bluck et al. (2023) we find that the stellar gravitational potential is predicted by several simulations to act as an excellent proxy for central black hole mass. In Section 4.1, we find that this remains true up to the earliest cosmic times (when simulations cease to predict the existence of quiescent systems). This provides a route to test the paradigm of AGN feedback quenching at early epochs, using photometric imaging data from JWST-CEERS, in addition to comparisons at lower redshifts with HST-CANDELS and the SDSS. Interestingly, recent papers have found that dynamically measured black hole mass does not correlate strongly with stellar mass in the early Universe, but is still closely connected to central velocity dispersion and, hence, the galaxy potential (see Maiolino et al. 2023; Paccuci et al. 2023). 

In Section 4.2 we find that the stellar potential, $\phi_*$, is unambiguously the best predictor of high mass galaxy quenching at very early cosmic times, exactly as predicted by simulations. Furthermore, we find that the predictive power of $\phi_*$ is completely stable over 13\,Gyr of cosmic history (see Fig. 3). This is in remarkable accord with predictions from the AGN feedback paradigm. Moreover, independent of the likely physical mechanism(s), this result demands a stable quenching process operating throughout cosmic time (see Bluck et al. 2022, 2023 for a similar conclusion at $z < 2$).

In Section 4.2, we also find that $\phi_*$ outperforms a host of other parameters in JWST-CEERS data, including stellar mass, galaxy size, galaxy 2D and 3D density, and Sersic index (amongst others). These results are naturally explained by the AGN feedback paradigm. Supermasisve black holes form and grow in dense potential wells (e.g. Sijacki et al. 2007, 2015; Weinberger et al. 2017). Once they reach a critical mass (typically $M_{BH} \sim 10^{7-8} M_{\odot}$; e.g., Weinberger et al. 2018), they transition to hot-mode accretion, form a stable central engine, and keep hot gas haloes stable from cooling and collapse for many Gyr. This process simultaneously resolves the cooling problem in galaxy groups and clusters (e.g., Fabian 2012); the cosmological low efficiency of star formation (e.g., Fukugita \& Peebles  2004); and the existence, and demographics, of quiescent galaxies (e.g., Peng et al. 2010, 2012; Bluck et al. 2014).

\subsubsection{Alternative Explanations to AGN}

\noindent We suspect that the AGN feedback paradigm is indeed the best explanation for the multi-epoch quenching results of the previous section. Certainly, all results we have found from JWST-CEERS, HST-CANDELS, and the SDSS are consistent with its predictions. Nevertheless, it is still worth considering if any other paradigm for massive galaxy quenching could also be consistent with our results.

An important early hypothesis for high mass galaxy quenching was to evoke galaxy - galaxy mergers. In such a scenario the ISM of the merging systems will be compressed leading to enhanced star formation, followed by a depression in star formation as a result of depleted gas reservoirs (e.g., Hopkins et al. 2006, 2008). Additionally, energetic feedback from quasar-mode AGN could aid in the depletion of the ISM by driving powerful outflows, ultimately triggering quenching. However, without some other form of long-term prevention of gas inflows (either from cold flows in low mass haloes, or from cooling of the CGM in high mass haloes) galaxies would inevitably rejuvenate, and no long-term quiescence would be apparent. This issue is brought into sharper focus by considering the `cooling catastrophe', whereby the CGM around massive galaxies, groups and clusters ought to be thermodynamically unstable, leading to runaway cooling, yet is observed to be stable across cosmic times (see, e.g., Fabian et al. 2006; Fabian 2012). As such, some heating source is required to maintain long term quiescence.

Given the logic of the cooling catastrophe (see Fabian 2012 for a review), a heating source is required to keep hot gaseous haloes buoyant, shielding galaxies from cold inflows (e.g., Dekel \& Birnboim 2006), and ultimately enabling quiescence through starvation of gas needed as fuel for future star formation (e.g., Peng et al. 2015; Bluck et al. 2020b). Astrophysically, there are a finite number of possibilities. The main three are: (i) AGN feedback (as discussed at length above, see Somerville \& Dav\'e 2015 for a review); (ii) stellar and supernova feedback (e.g., Cole et al. 2000; Henriques et al. 2015); and (iii) virial shocks (e.g., Dekel \& Birnboim 2006; Woo et al. 2013). Additionally, more exotic alternatives could be speculated to be important, such as dark matter annihilation, magnetic field heating, and cosmic rays. However, the latter three are not well calibrated, have weak-to-no observational evidence, or are not predicted by cosmological models to impact the star formation efficiency problem. As such, we restrict the following discussion to the three numbered sources above.

In Bluck et al. (2020a) we demonstrate that supernova (or more generally stellar) feedback must scale fundamentally with stellar mass, as the time integral of star formation rate. In Section 4.2 (see also Bluck et al. 2014; 2016; 2020a,b, 2023; Piotrowska et al. 2022; Brownson et al. 2022) we find that stellar mass is of no intrinsic importance for galaxy quenching at any epoch. Indeed, the importance of stellar mass tends to that of a random variable, once more fundamental parameters are included (see Figs. 3 \& 4). This emphatically rules out supernovae as the fundamental cause of high mass galaxy quiescence. Indeed, this result is entirely as expected in modern cosmological models, which introduced AGN feedback explicitly to deal with the failure of supernovae to quench galaxies (see Croton et al. 2006; Bower et al. 2006, 2008; Sijacki et al. 2007). 

Also in Bluck et al. (2020a), we show that the heating induced by virial shocks scales as a power law with halo mass (consistent with the presentation in Dekel \& Birnboim 2006). Hence, halo mass is the key observable of virial shock quenching, which is why it has been dubbed `halo mass quenching' (see Woo et al. 2013, 2015). Although we do not have access to reliable estimates of halo masses in the early Universe yet, at lower redshifts we have found that halo mass is of no greater predictive power over quenching than a random variable, once an estimate of black hole mass is included (see Bluck et al. 2016, 2020a, 2022; Piotrowska et al. 2022). This rules out halo mass quenching in the local Universe. Moreover, the extreme stability of the photometric parameters' quenching importances (see Fig. 3) strongly indicates that there is no variation in quenching mechanism across cosmic time. Hence, we {\it predict} that halo mass will not be important at high redshifts, exactly as in the local Universe. This prediction may be tested relatively soon, using abundance matching in the wide-field spectroscopic VLT-MOONRISE survey (Maiolino et al. 2020).

One final possibility is that the strong dependence of quenching on the galactic potential (traced by $\phi_*$) in observations could be a result of dynamical stabilization (e.g., Martig et al. 2009; Gensior et al. 2020). In such a scenario, dynamical torques from a central mass distribution could provide stabilization against gravitational collapse of the ISM in disk structures. Consequently, the gas content of galaxies is predicted to {\it rise} during quenching because in quiescent systems gas conversion into stars halts, yet gas accretion into the system (either from cold gas accretion or CGM cooling) continues. However, in the local Universe the opposite is seen, whereby quiescent systems have significantly lower gas fractions than their star forming counterparts (e.g., Saintonge et al. 2017; 2018; Piotrowska et al. 2020; Brownson et al. 2020; Ellison et al. 2020; 2021a,b). Hence, this cannot be the dominant quenching mechanism at low redshifts. 

At higher redshifts it remains to be seen whether the gas content of quiescent objects are lower than star forming systems. However, long-term quiescence cannot be maintained without shutting off gas inflow into the system (as demonstrated by simulations, see, e.g., Vogelsberger et al. 2014a,b; Schaye et al. 2015; Nelson et al. 2018). No amount of dynamical stabilization could prevent the vast quantities of gas accreted into galaxies from forming stars, if cooling from the CGM is not prevented (see Fabian 2012 for a discussion). Hence, we reject the dynamical stabilization possibility as a probable quenching mechanism for high-mass galaxies across cosmic time.

We remain agnostic about the possibility of future proposed heating sources driving galaxy quiescence. However, any such mechanism must simultaneously explain: (i) why $\phi_*$ is the best photometric predictor of quiescence; (ii) why this feature is completely stable across cosmic time; and (iii) why the observed long term heating from AGN (e.g., Fabian 2012; Hlavacek-Llarondo et al. 2015, 2017, 2018), which is consistent with the two criteria above, does not stabilize the CGM around massive galaxies and hence requires another quenching source. Applying Occam's razor, it seems that the AGN feedback paradigm is the most probable quenching mechanism for massive galaxies across the entire history of the Universe. Furthermore, this conclusion is in perfect accord with the current (near consensus) theoretical view.

\subsection{Environmental Quenching in the Early Universe}

\noindent In this paper we have focused primarily on massive galaxy quenching ($M_* > 10^{9.5} M_\odot$ systems). As such, we have sought to test the AGN feedback paradigm across cosmic time, with the main novelty being the high and ultra-high redshift regimes accessible for the first time with JWST-CEERS. However, in Section 4.2.3 we expand our view of galaxy quiescence to lower stellar masses and, hence, are sensitive to other quenching mechanisms.

In Figs. 6 \& 7 we see two distinct populations of quiescent galaxies, both existing at all epochs studied (from $z= 0.5 - 8$). For high mass galaxies ($M_* > 10^{10} M_\odot$), we see clear evidence of quenched systems having higher $\phi_*$, at fixed $M_*$, than their star forming counterparts. This is completely consistent with the AGN feedback paradigm (see Section 5.1). Additionally, we find a population of quenched low mass galaxies ($M_* < 10^{10} M_\odot$) which are {\it not} offset in $\phi_*$, at fixed $M_*$. Moreover, since these systems are defined to have low stellar masses, this population of quiescent galaxies have their quenching uncorrelated with both stellar mass and central potential. Consequently, this type of quenching is likely uncorrelated with black hole mass and, hence, AGN feedback.

In the local Universe, a similar population of low mass (and shallow gravitational potential) quiescent objects are identified, which are found to have their quiescence critically connected to environment (e.g., Peng et al. 2010, 2012; Bluck et al. 2014, 2016; Goubert et al. in prep.). An analogous population is also identified out to $z \sim 2$ by, e.g., Lang et al. (2014). As such, it is logical to speculate that the low mass quenched galaxies observed out to $z \sim 8$ in JWST-CEERS are similarly environmentally quenched systems. We also note that the simulations used in this work (Eagle and TNG) both predict the existence of low mass quenched systems, uncorrelated with black hole mass, which are predominantly satellites (see Goubert et al. in prep. for an analysis of satellite quenching at early cosmic times).

At present, it is not at all straightforward to quantify the environments of JWST-CEERS galaxies. The primary limitation is the lack of spectroscopic redshifts for a large contiguous galaxy sample at high-$z$, which is required for accurate group and cluster identification. In the coming $\sim$5 years the lack of wide-field spectroscopic surveys targeting the early Universe will be addressed with VLT-MOONRISE (Maiolino et al. 2020), as well as various planned JWST spectroscopic surveys. As such, we take the opportunity to make a prediction: The observed low-mass quenched galaxies in JWST-CEERS will turn out to be environmentally quenched objects, as evidenced by a strong correlations between their star formation and galaxy environmental metrics (such as local density, central - satellite categorization, and halo properties).

Using the rationale above, we suspect that we have found (indirect) evidence for environmental quenching at the earliest cosmic times ever probed. Testing this hypothesis will likely become a high priority of the next generation of high-$z$ spectroscopic galaxy surveys.

\section{Summary}

\noindent In this paper we start by extracting the testable predictions for AGN feedback driven quenching at very high redshifts from the EAGLE and TNG cosmological simulations. We then focus on testing these predictions using data from JWST-CEERS  data in the very early Universe, complementing this analysis with observational tests at later cosmic epochs utilizing HST-CANDELS and the SDSS. 

Throughout the paper we use a consistent definition of quiescence in galaxies, defined as sSFR values lower than the inverse of the Hubble time at each redshift (see Fig.~1). We also test the stability of our results against a more stringent quenching threshold (a factor of three lower than this). The only exception to this general approach is at $z \sim 0$ (probed by the SDSS), where the inverse Hubble time ceases to be a viable threshold (see Bluck et al. 2023). In this case we adopt the commonly used threshold at sSFR $< 10^{-11}\,{\rm yr}^{-1}$ (as in Piotrowska et al. 2022; Bluck et al. 2023).

To analyse both the observational and simulated galaxy surveys, we utilize a sophisticated statistical approach with Random Forest classifications, segregating star forming and quiescent galaxies. We have demonstrated that our Random Forest classification technique is extremely effective at identifying causal relationships in complex inter-correlated data (see Bluck et al. 2022; Piotrowska et al. 2022; Brownson et al. 2022).\\

Our primary results are as follows:\\

I) Both TNG and EAGLE predict that supermassive black hole mass should be the key predictor of quiescence in the early Universe at $z = 2 - 4$ (see Fig.~2, top panel). This is identical to the predictions from these simulations at lower redshifts, which is indicative of a single quenching mechanism operating throughout the history of the Universe in these models (see Piotrowska et al. 2022; Bluck et al. 2023 for results at $z < 2$). \\

II) Both TNG and EAGLE also predict that, in lieu of supermassive black hole mass, the stellar potential ($\phi_* = M_* / R_h$) will outperform stellar mass, and several other parameters, in predicting quenching (see Fig.~2, bottom panel).\\

III) In JWST-CEERS, at both intermediate ($0.5 < z < 2$) and high ($2 < z < 4$) redshifts, we find that $\phi_*$ is unambiguously identified as the best parameter for predicting quiescence in galaxies, from a Random Forest classification. Once $\phi_*$ is accounted for, there is essentially no importance for quenching contained in the stellar mass, galaxy size, stellar surface (or volume) density, or several other galaxy structural parameters (see Figs.~3 - 5).\\

IV) Through comparison with HST-CANDELS and the SDSS, we find that the key parameter for predicting high mass galaxy quiescence is stable across 13\,Gyr of cosmic history. This indicates a very high probability of a single high-mass galaxy quenching mechanism operating throughout all of cosmic time (see Fig.~3).\\

V) We identify two distinct populations of quiescent galaxies at high redshifts in JWST-CEERS: (i) high mass quenched galaxies, which have significant offsets to higher $\phi_*$ at a fixed $M_*$ (as expected in the AGN feedback paradigm); and (ii) low mass quenched galaxies, which have no offsets in $\phi_*$ (see Figs.~6 \& 7). Through analogy with low-$z$ observations, we speculate that the latter population is likely a result of environmental quenching in the early Universe (see Section 5.2).\\

VI) Finally, we find that, even though the importance of $\phi_*$ for predicting quenching remains stable across cosmic time, the $f_Q - \phi_*$ relation does evolve, such that higher values of $\phi_*$ are required in order for galaxies to quench at earlier cosmic times (see Figs.~8 \& 9).\\

Taken as a whole, the observational results from this paper represent a critical test of the AGN feedback paradigm, spanning over 13\,Gyr of cosmic history. The direct predictions from this paradigm (as given by cosmological simulations) are in remarkable agreement with the observations. This strongly suggests that AGN feedback can explain high mass galaxy quenching at all epochs in the history of the Universe. We emphasize that a more `direct' test utilising a putative correlation between sSFR and AGN luminosity is {\it not} predicted to exist by contemporary AGN feedback models (see Section 5.1 for a discussion on this crucial point). Hence, the methodology of this work is essentially the optimal strategy to test the AGN feedback paradigm across cosmic time (at least in photometric data).

We also identify the need for low-mass galaxy quenching in JWST-CEERS, which is incompatible with AGN feedback (since it occurs in galaxies which, almost certainly, host low mass central black holes). In the local Universe, low mass quenching is near exclusively associated with environment. Hence, our high-$z$ results suggest that environment may already suppress star formation for some low mass systems as early as 1 - 2\,Gyr after the Big Bang. Further exploration of this population will require wide-field spectroscopic galaxy surveys targeting the early Universe, which will be realized in the next $\sim$5 years with VLT-MOONRISE, and various upcoming JWST surveys.

\section*{Data Access}

\noindent All of the data used in this paper is publicly available, or else currently in preparation for public release.\\

\noindent {\it HST+JWST CEERS Data}:\\
For access to photometric redshift and stellar mass catalogs see Duncan et al. (2019). \\
For access to structural catalogs see Ormerod et al. (2023, in prep.). Or, for immediate requirements, message the corresponding author with reasonable requests. \\
Full details on the JWST observations and source data are provided in: https://ceers.github.io/ \\

\noindent {\it HST-CANDELS}:\\
Source data: https://archive.stsci.edu/hlsp/candels \\
VAC: https://mhuertascompany.weebly.com/data-releases- and-codes.html \\

\noindent {\it SDSS}:\\
Source data: https://classic.sdss.org/dr7/access/ \\
SFR catalog: See Brinchmann et al. (2004). \\
Morphological catalog: See Simard et al. (2011)\\
Stellar mass catalog: See Mendel et al. (2014). \\

\noindent {\it Simulations}:\\
Eagle data access: http://icc.dur.ac.uk/Eagle/ \\
IllustrisTNG data access: www.tng-project.org/ \\
Docker: https://hub.docker.com/u/jpiotrowska\\

\noindent {\it Analysis}:\\
All analyses in this work were performed using {\small PYTHON-3} (https://www.python.org/), including {\small NUMPY, ASTROPY, SCIPY, PANDAS, SEABORN, MATPLOTLIB}. All machine learning analyses were performed using {\small SCIKIT-LEARN} (Pedregosa et al. 2011; see: https://scikit-learn.org).

\section*{Acknowledgements}

\noindent We thank the anonymous referee for a constructive report, which has helped us to improve the presentation of our work. We thank all contributors to JWST-CEERS, HST-CANDELS, the SDSS, the Eagle collaboration, and the Illustris and IllustrisTNG collaborations. This paper would not have been possible without much work done by hundreds of scientists, engineers, and support staff over the past decade. We also thank the teams behind the analysis tools we have used from {\small PYTHON}, especially those involved in {\small SCIKIT-LEARN}.

AFLB gratefully acknowledges a faculty start-up grant at the Florida International University. CJC acknowledges support from the ERC Advanced Investigator Grant EPOCHS (788113), as well as studentships from STFC. NJA acknowledges support from the ERC Advanced Investigator Grant EPOCHS (788113). D.A. and T. H. acknowledge the support of studentships from the STFC. KJD acknowledges support from the STFC through an Ernest Rutherford Fellowship (grant number ST/W003120/1). RM acknowledges a Royal Society Research Professorship, as well as support from the STFC and ERC Advanced Grant (695671) `QUENCH'.

\bibliography{sample63}{}
\bibliographystyle{aasjournal}

\appendix

\section{Machine Learning Reproducibility}

\begin{table}
\begin{center}
\caption{RF Hyperparameters for Fig. 2}
\begin{tabular}{ c c c c c c c c c c c  } 
 \hline
  & Method & $N_{\rm trees}$ & Max Depth & $N_{\rm iterations}$  & Balanced & Normalized & Train : Test & MSL & AUC & $\Delta$AUC \\ 
  \hline
  \hline
Eagle (ALL) & RFclass  & 250  & 250  & 10 & Yes  & Yes  & 70 : 30  & 60  & 0.95  & 0.02   \\ 
Eagle (Obs) & RFclass  & 250  & 250  & 10 & Yes  & Yes  & 70 : 30  & 60  & 0.86  & 0.01   \\ 
IllustrisTNG (ALL) & RFclass  & 250  & 250  & 10 & Yes  & Yes  & 70 : 30  & 25  & 0.98  & 0.02   \\ 
IllustrisTNG (Obs) & RFclass  & 250  & 250  & 10 & Yes  & Yes  & 70 : 30  & 25  & 0.95  & 0.02   \\ 
 \hline
\end{tabular}
\end{center}
\end{table}

\begin{table}
\begin{center}
\caption{RF Hyperparameters for Fig. 3}
\begin{tabular}{ c c c c c c c c c c c  } 
 \hline
  & Method & $N_{\rm trees}$ & Max Depth & $N_{\rm iterations}$  & Balanced & Normalized & Train : Test & MSL & AUC & $\Delta$AUC \\ 
  \hline
  \hline
JWST ($2 < z < 4$) & RFclass  & 250  & 250  & 100 & Yes  & Yes  & 70 : 30  & 30  & 0.87  & 0.02   \\ 
JWST ($0.5 < z < 2$) & RFclass  & 250  & 250  & 100 & Yes  & Yes  & 70 : 30  & 50  & 0.79  & 0.02   \\ 
HST-CANDELS & RFclass  & 250  & 250  & 25 & Yes  & Yes  & 50 : 50  & 110  & 0.78  & 0.02   \\ 
SDSS & RFclass  & 250  & 250  & 10 & Yes  & Yes & 50 : 50  & 1000  & 0.82  & 0.02   \\ 
 \hline
\end{tabular}
\end{center}
\end{table}

\begin{table}
\begin{center}
\caption{RF Hyperparameters for Fig. 4}
\begin{tabular}{ c c c c c c c c c c c  } 
 \hline
  & Method & $N_{\rm trees}$ & Max Depth & $N_{\rm iterations}$  & Balanced & Normalized & Train : Test & MSL & $<{\rm AUC}>$ & $<\Delta{\rm AUC}>$ \\ 
  \hline
  \hline
JWST ($\lambda$) & RFclass  & 250  & 250  & 100 & Yes  & Yes  & 70 : 30  & 65  & 0.79  & 0.02   \\ 
 \hline
\end{tabular}
\end{center}
\end{table}

\noindent In order to enable accurate reproduction of our machine learning results by other researchers, we provide in this appendix a series of tables listing the hyperparameters used in the RF classification analyses (for Figs. 2 - 4 see Tables 2 - 4, respectively). 

The table rows indicate each classification run, which all categorize star forming and quiescent galaxies via a threshold set at the inverse Hubble time (or sSFR $< 10^{-11} {\rm yr}^{-1}$ for the SDSS at $z=0.1$). All classifications are run using {\small SCIKIT-LEARN} (Pedregosa et al. 2011; see https://scikit-learn.org)

In every classification run we set {\it max\_features} = `None'. This ensures that every feature is available at every node in each decision tree within the Random Forest. In Bluck et al. (2022) we have demonstrated that this mode represents the most effective RF classification architecture for disentangling inter-correlated `nuisance' parameters from the underlying causal relations in simple models, simulations, and observations. Note that this is {\it not} the default method of RF classification in {\small SCIKIT-LEARN}. However, it is the most effective for revealing physical insight.\\

\noindent Table columns are as follows:\\

\noindent {\it Method: } Type of machine learning used. Here always RandomForestClassifier from {\small SCIKIT-LEARN} (abbreviated to RFclass in the tables).\\

\noindent {\it $N_{\rm trees}$: } Number of decision tress used in the random forest. Here set universally to 250. We have tested that increasing the number of trees does not further improve the results of any of these classifications.\\

\noindent {\it Max Depth}: Number of decision forks permitted within each decision tree. This is set universally to 250, which is deeper than ever required (given the MSL limit, see below).\\

\noindent {\it $N_{\rm iterations}$: } The number of times we run each entire RF classification, using different randomly selected subsets of the data for training and testing. These iterations are used to infer statistical uncertainties on the feature importances. Note that this number varies from 10 - 100, to account for very different data sizes (and hence run times and sensitivity to random sampling). \\

\noindent {\it Balanced: } Whether the training and validation data has an equal number of star forming and quenched galaxies within it. We follow standard best practice by evenly sampling each class in every RF classification run (see Teimoorinia et al. 2016 for a discussion).\\

\noindent {\it Normalized: } Whether the training data is normalized prior to being given to the classifier. In this work we always median subtract, and normalize by the inter-quartile range, every feature in the classifier. This ensures a fair comparison between data with different numerical values and distribution shapes (see Bluck et al. 2022 for further discussion).\\

\noindent {\it Train : Test:} The ratio of data sizes used in training and testing. In general a 50 : 50 split is ideal. However, for smaller data sets it is recommended to weight more heavily on training. As such, we adopt 70 : 30 for some of the runs. We have tested that this leads to more stable results (lower errors) and higher overall performance in those cases.\\

\noindent {\it Min-Samples-Leaf (MSL): } This hyperparameter sets the minimum number of data in a given node of a given decision tree needed to attempt further splits in the data. Setting this value to its logical minimum of two enables perfect classification of the training data. However, this leads to overfitting. We adjust this parameter to simultaneously maximize the performance in training (AUC, penultimate column) whilst maintaining a minimal difference in performance with unseen testing data ($\Delta$AUC, final column). Explicitly, in this work we require  $|\Delta {\rm AUC}| < 0.02$, yielding extremely similar performance in unseen data to the training sample, thus solving the overfitting problem. \\

\noindent {\it AUC: } The area under the true positive - false positive receiver opertor curve, which quantifies the overall performance of the RF classification (see Teimoorinia et al. 2016). A value of AUC = 0.5 indicates a pure random performance, whereas a value of AUC = 1 indicates a perfect segregation of star forming and quiescent systems. Note that in the table the AUC is given for training data.\\

\noindent {\it $\Delta$AUC: } The difference in AUC between training and testing data. Following Bluck et al. (2022), we adopt a maximum threshold of 0.02 to count as a successful fit, i.e. free of significant overfitting.\\

We emphasize that the parameters shown in the tables indicate the optimal settings for each classification run. In addition to these runs we have explored a wide hyperparameter space, involving significant variation in essentially all of the variables discussed above. All results in this paper are extremely stable to these alternative modes of classification, within reasonable bounds. As such, the results presented in the main body of the paper are easily obtainable without carefully matching our prescriptions. Nonetheless, to fully reproduce our results the above settings should be used.

\newpage

\section{Predicted $M_{BH} - \phi_*$ Relation in Simulations at Early Cosmic Times}


\begin{figure*}
\begin{centering}
\includegraphics[width=0.49\textwidth]{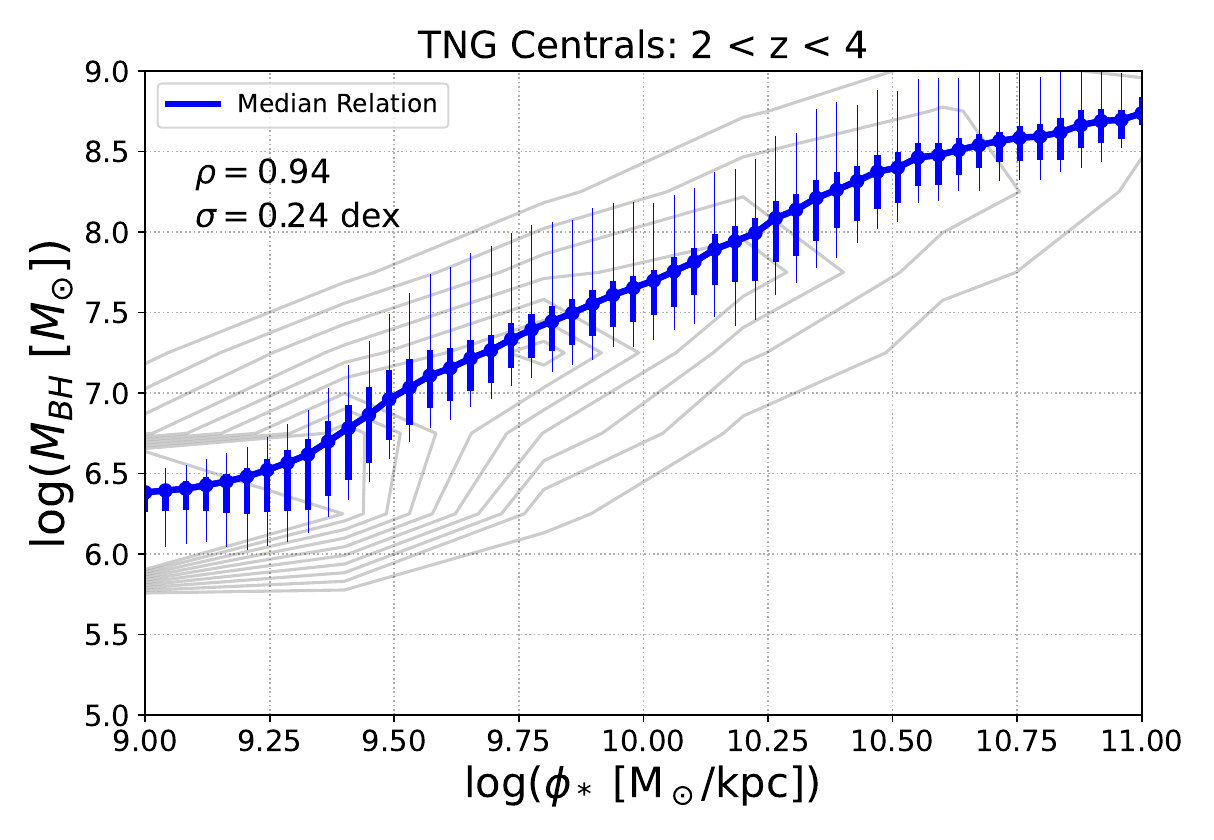}
\includegraphics[width=0.49\textwidth]{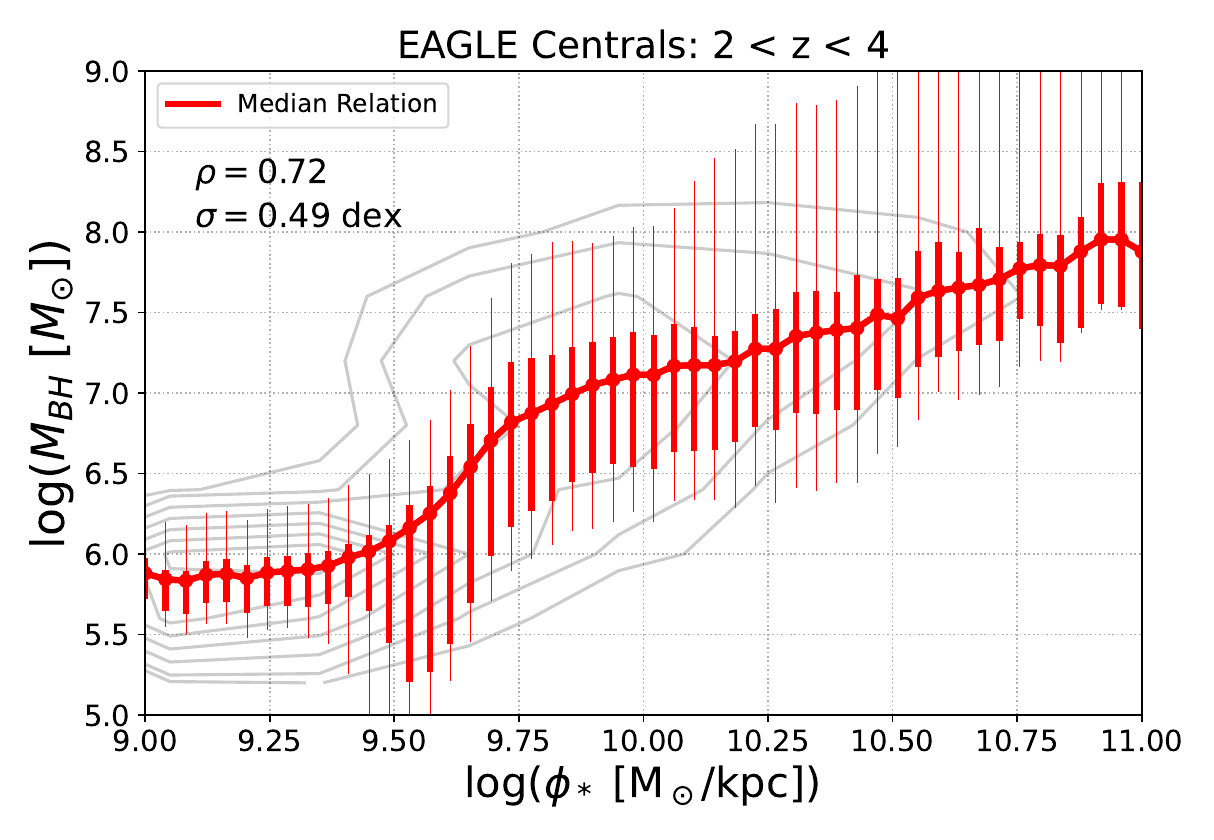}
\caption{The relationship between supermassive black hole mass and stellar gravitational potential in simulations of the early Universe, shown separately for TNG (left) and EAGLE (right). The solid lines indicate the median relationship, with solid error bars indicating the 1$\sigma$ dispersion, and light error bars indicating the 2$\sigma$ dispersion. Linearly spaced density contours are overlaid in grey. Additionally, on each panel we present the Spearman rank correlation coefficient ($\rho$), and the average galaxy weighted dispersion from the median relation ($\sigma$). It is clear that both simulations predict a strong relationship between $M_{BH}$ and $\phi_*$, which is leveraged in this work to test the integrated effect of AGN feedback at early cosmic times in photometric data. However, note that this relationship exhibits a stronger correlation and tighter dispersion in TNG compared to EAGLE.}
\end{centering}
\end{figure*}

\noindent In Section 4.1 we show that the fundamental parameter regulating central galaxy quenching in cosmological simulations is supermassive black hole mass (see Fig. 2, upper panel). However, we do not have access to black hole masses for galaxies in the early Universe for the photometric surveys used in this work. Furthermore, accurate dynamical measurements for black hole masses will likely only ever become available for less than 10\% of galaxies at these epochs (i.e., for systems which present as powerful AGN with the appropriate orientation of molecular torus to measure broad line regions). Hence, dynamical measurements of black hole masses for early galaxies will never yield a fully representative sample, which is needed for statistical analysis of galaxy demographics (e.g., star forming vs. quiescent populations, as studied in this paper).

To combat this issue, we have looked for a proxy for black hole mass to use in extant photometric data, in order to test the AGN feedback paradigm from simulations at early cosmic times. In the lower panel of Fig. 2, we find that in lieu of a measurement of black hole mass, the stellar potential ($\phi_*$) is predicted to become the most effective predictor of quenching in simulations. This is a clear {\it prediction} of the simulations in the a-causal photometric parameter space. The primary goal of the rest of the paper has been to test this prediction across cosmic time in JWST-CEERS, HST-CANDELS, and the SDSS. This expands of previous work in Piotrowska et al. (2022) and Bluck et al. (2023).

The reason $\phi_*$ works as a good proxy for $M_{BH}$ is that the two parameters are strongly correlated. In Fig. 10 we show the $M_{BH} - \phi_*$ relationship for TNG (left) and EAGLE (right) at early cosmic times ($2 < z < 4$). See Bluck et al. (2023) for a presentation of these relationships at later cosmic epochs. In both simulations, there is a strong correlation between the two parameters, with a relatively tight scatter (see statistics presented on each panel). In fact the scatter on this relationship is comparable, or tighter, to that found between black hole mass and central velocity dispersion in the local Universe (see Saglia et al. 2016; Terrazas et al. 2016, 2017; Piotrowska et al. 2022). We have tested that this relationship is the most effective to use in photometric data to constrain supermassive black hole masses via a variety of techniques. Most importantly, in the lower panel of Fig. 2 we find that, in lieu of a measurement of black hole mass, the stellar potential becomes the most predictive observable parameter of quiescence. In essence, Fig. 10 shows why this works.

In the simulations, the origin of these strong relationships is due to supermassive black holes forming and growing preferentially in deep potential wells. See Bluck et al. (2023) for a discussion. Note that the correlations between these parameters are not seen explicitly in the Random Forest analysis because when supermassive black hole mass is controlled for (as in the top panel of Fig. 2), there is no residual dependence of quenching on the stellar potential. This exposes the stellar potential's correlation with quenching as being spurious. Nonetheless, the stellar potential is still a highly useful parameter because we can leverage it as a proxy for black hole mass in photometric data, much as the central velocity dispersion is often used as a proxy for black hole mass in spectroscopic data (see, e.g., Bluck et al. 2022; Brownson et al. 2022; Piotrowska et al. 2022).

\end{document}